# The logic of interactive Turing reduction

Giorgi Japaridze[*]

Department of Computing Sciences, Villanova University
800 Lancaster Avenue, Villanova, PA 19085, USA
Email: giorgi.japaridze@villanova.edu
URL: http://www.csc.villanova.edu/~japaridz/

**Abstract**

The paper gives a soundness and completeness proof for the implicative fragment of intuitionistic calculus with respect to the semantics of computability logic, which understands intuitionistic implication as interactive algorithmic reduction. This concept — more precisely, the associated concept of reducibility — is a generalization of Turing reducibility from the traditional, input/output sorts of problems to computational tasks of arbitrary degrees of interactivity.

## 1 Introduction

A quest for a convincing semantical explication of the constructivistic philosophy of intuitionistic logic has been on for many decades. Among the most notable but unsuccessful attempts in this direction are Kleene's realizability [12], Gödel's Dialectica interpretation [5] and Medvedev's finite-problem semantics [16], directly or indirectly stimulated by Kolmogorov's [14] well known yet rather abstract thesis, according to which *intuitionistic logic is a logic of problems*. A series of semantics in this style, while meaningful and close in spirit to Kolmogorov's vision, turned out to yield inherent incompleteness. Blass's appealing game semantics [1] which, by the way, happens to be the nearest precursor of the semantics dealt with in the present paper,[1] should also be listed here. While the propositional fragment of intuitionistic logic was proven to be sound with respect to it, the questions on predicate-level soundness and whatever-level completeness have been left open for posterity. As for Lorenzen's game semantics [3, 15], it has been seen as merely a technical servant to the existing syntactic construction rather than meaningful and inspiring enough in its own right to provide an ultimate justification for intuitionistic logic.

---

[*]This material is based upon work supported by the National Science Foundation under Grant No. 0208816, and 2005 Summer Research Grant from Villanova University.

[1]A detailed discussion of how the two semantics compare is given in Sections 13 and 27 of [6].

In [6] the author introduced a program and framework for redeveloping logic as a formal theory of computability, as opposed to the formal theory of truth which it has more traditionally been. At the heart of that approach, baptized as *computability logic* (CL), is a game semantics. Games in CL are seen as formal equivalents of our broadest intuition of computational problems, specifically, computational problems in their most general — interactive — sense. Computability, or algorithmic solvability, of such problems is understood as existence of an interactive Turing machine that wins the game no matter how the other player, which is the environment or the user, behaves. In this vision of logic, computability replaces the classical concept of truth, operations on games=problems replace Boolean connectives and quantifiers, and the understanding of logical formulas as (representing) computational problems comes in the place of their classical understanding as true/false statements, with valid formulas now describing universally correct principles of computability, and the logic in whole providing a systematic answer to the question on what and how can be algorithmically solved.

This characterization makes the qualification "*logic of problems*" adequate for CL. Hence, in a search for a direct materialization of Kolmogorov's thesis and (thus) the constructivistic claims of intuitionistic logic, the semantics of CL should be a reasonable candidate to try. Indeed, while the ambition of CL is to be a good tool for successfully navigating the real — non-man-made — world and, accordingly, its semantics has been created with no intention whatsoever to serve or justify any existing deductive systems, at the very time of its conception the conjecture was born[2] that the set of theorems of Heyting's intuitionistic calculus is precisely what one gets by restricting the otherwise more expressive CL (here identified with the set of its semantically valid formulas) to a certain natural sublanguage, called in [6] the *intuitionistic fragment*. It should be noted that, unlike intuitionistic or linear logics, CL is a semantically rather than syntactically introduced theory and, at present, many questions on possible axiomatizations of various fragments of it, including the intuitionistic fragment, being far from trivial, remain open.

The first step toward a positive verification of the above conjecture has been made in [11], where the soundness of the full first-order intuitionistic calculus with respect to the CL semantics was proven. As always, however, finding a completeness proof — provided that one exists, of course — presents a greater challenge. The right strategy could be to approach the problem in a patient, step-by-step manner, starting from some relatively simple yet nontrivial fragment of the logic. The present paper tells a first success story on the way of following that strategy. It gives a completeness proof for the implicative fragment of propositional intuitionistic calculus.

We are going to see that there are two operations of computability logic, $\succ\!\!-$ and $\circ\!\!-$, that behave intuitionistically.[3] That is, the implicative fragment of intuitionistic logic is sound and complete regardless whether its implication is understood as $\succ\!\!-$ or $\circ\!\!-$. The "real" intuitionistic implication, however, is $\circ\!\!-$ rather than $\succ\!\!-$. That is because the behavior of $\succ\!\!-$ is exactly intuitionistic only in isolation, and stops being so when $\succ\!\!-$ is taken in combination with other operators: the full intuitionistic calculus with $\succ\!\!-$ as implication is simply unsound, even though it most likely remains complete. For example, one could show that the intuitionistically provable formula

$$(P \succ\!\!- R) \succ\!\!- \Big((Q \succ\!\!- R) \succ\!\!- \big((P \sqcup Q) \succ\!\!- R\big)\Big),$$

where $\sqcup$ means disjunction, fails to be a valid principle of computability logic. The official conjecture of [6] regarding the soundness and completeness of the full intuitionistic logic, and the verification of its soundness part in [11], assume reading intuitionistic implication as $\circ\!\!-$, and reading intuitionistic conjunction, disjunction and quantifiers as what computability logic calls *choice operations* $\sqcap, \sqcup, \sqcap\!\!\!\sqcap, \sqcup\!\!\!\sqcup$.

---

[2]Officially stated in [6] as Conjecture 25.8.

[3]The operation $\circ\!\!-$ in [6] was denoted by $\Rightarrow$; it was renamed into $\circ\!\!-$ later in [9, 10]. In this paper $\Rightarrow$ has the different meaning of that of a separator of the two parts of a sequent.

Choice operations in CL model decision steps in the course of interaction, with positive occurrences of disjunction and existential quantifier meaning machine's choices, and positive occurrences of conjunction and universal quantifier meaning choices by its environment (in negative occurrences, the roles are interchanged). For instance, where $f(x)$ is a function, $\sqcap\!\!\!\sqcap x \sqcup\!\!\!\sqcup y (y = f(x))$ is the game in which the first move/action/choice is by the environment, consisting in specifying a particular value $m$ for $x$. Such a move, which intuitively can be seen as asking the machine the question "*what is the value of $f(m)$?*" brings the game down to the position $\sqcup\!\!\!\sqcup y(y = f(m))$. The next step is by the machine, who should specify a value $n$ for $y$, further bringing the game down to $n = f(m)$. The latter is understood as a moveless position won by the machine if true and lost if false.[4] The machine's move $n$ can thus be seen as answering/claiming that $n$ is the value of $f(m)$. From this explanation it must be clear that $\sqcap\!\!\!\sqcap x \sqcup\!\!\!\sqcup y (y = f(x))$ represents the problem of computing $f$, with the machine having an algorithmic winning strategy for this game iff $f$ is a computable function. Similarly, where $S$ is a set, $\sqcap\!\!\!\sqcap x(x \in S \sqcup x \notin S)$ represents the problem of deciding $S$: here, again, the first move is by the environment, consisting in choosing a value $m$ for $x$ (asking "*is $m$ an element of $S$?*"); and the next step is by the machine who, in order to win, should choose the true disjunct of $m \in S \sqcup m \notin S$, i.e. correctly tell whether $m$ is an element of $S$ or not.

As for $\succ\!\!-$ and $\circ\!\!-$, these are two flavors of the operation of *reducing* one problem to another. Either one, as a reduction operation, induces — and will be often understood as — a reducibility relation. Specifically, a problem $B$ is $\succ\!\!-$-*reducible* or $\circ\!\!-$-*reducible* to a problem $A$ iff the corresponding compound problem $A \succ\!\!- B$ or $A \circ\!\!- B$ has an algorithmic solution. Both versions of reducibility are natural and important as both present conservative extensions of Turing reducibility. That is in the sense that they are equivalent to Turing reducibility when restricted to the traditional, two-step, input/output sorts of problems. $\succ\!\!-$ and $\circ\!\!-$ start to diverge only when their scope is extended to problems with higher degrees of interactivity — the kind of problems to which Turing reducibility is inapplicable for the simple reason of never having been defined or generalized.

---

[4]This should not suggest that the atoms of intuitionistic logic are required to be interpreted as moveless games/positions. Rather, they represent arbitrary interactive computational problems.

Intuitively, both $A \succ\!\!- B$ and $A \circ\!\!- B$ are problems solving which means solving $B$ while having $A$ as an external computational resource, with an "external computational resource" meaning a to-be-solved-by-the-environment computational problem. More precisely, in the antecedent of an implication, the roles of the problem-solving machine and its environment are interchanged. Acting in the role of an environment there, the machine may observe how the antecedent is being solved, and utilize this information in its own solving the consequent.

An example might help. Let $T$ be a finitely axiomatized applied theory based on classical first-order logic, such as, say, Robinson's arithmetic, and let $x$ range over the formulas of the language of $T$. Next, let $Pr_T(x)$ be the predicate "$x$ is a theorem of $T$", $Pr(x)$ the predicate "$x$ is provable in classical predicate calculus", and $\neg Pr_T(x)$ and $\neg Pr(x)$ the ordinary negations of these predicates. As a decision problem, the $T$-theoremhood problem would be expressed by $\sqcap x\bigl(Pr_T(x) \sqcup \neg Pr_T(x)\bigr)$. This is generally undecidable, yet algorithmically reducible to the problem $\sqcap x\bigl(Pr(x) \sqcup \neg Pr(x)\bigr)$ of provability in predicate calculus. The problem of reducing the former to the latter can then be expressed by

$$\sqcap x\bigl(Pr(x) \sqcup \neg Pr(x)\bigr) \quad \supset \quad \sqcap x\bigl(Pr_T(x) \sqcup \neg Pr_T(x)\bigr), \tag{1}$$

where $\supset$ is either one of the operators $\succ\!\!-$, $\circ\!\!-$. The obligation of a machine solving the above compound problem is to solve $\sqcap x\bigl(Pr_T(x) \sqcup \neg Pr_T(x)\bigr)$, i.e. answer any given question of the type "*is $F$ a theorem of $T$?*" asked by the environment. Yet, the machine is expected to do so only on the condition that the environment does not fail to correctly solve the similar problem $\sqcap x\bigl(Pr(x) \sqcup \neg Pr(x)\bigr)$ in the antecedent, where the roles are switched and it is the machine who can ask questions like "*is $H$ provable in predicate calculus?*". So, here is an algorithmic strategy for the machine. Wait until the environment asks "*is $F$ a theorem of $T$?*" for some particular $F$. Then ask the counterquestion "*is $Ax \to F$ provable in predicate calculus?*", where $Ax$ is the conjunction of all non-logical axioms of $T$. The environment will have to provide a correct yes/no answer, or else it loses. Whatever the answer of the environment in the antecedent is, repeat the same answer in the consequent, and rest your case. A success of this strategy is guaranteed by the deduction theorem for classical logic.

Observe that the above explanation of the meaning of $A \supset B$ (where $\supset \in \{\succ\!\!-, \circ\!\!-\}$) is exactly what the Turing reduction of $B$ to $A$ is all about. The latter — more precisely, the *Turing reducibility* of $B$ to $A$ — is defined as existence of a Turing machine that solves $B$ when having an oracle correctly answering any questions regarding $A$. We can see such an oracle as (a part of) the environment solving $A$ for the machine, thus providing an external computational resource. The only difference is that the resource provided by an oracle is always a simple question-answering type of a task such as the above $\sqcap x\bigl(Pr(x) \sqcup \neg Pr(x)\bigr)$, while the antecedent (as well as the consequent, of course) of a $\succ\!\!-$- or $\circ\!\!-$-implication can have an arbitrarily complex interaction interface. For instance, (1) is already a problem with a non-standard interface, and CL, unlike traditional approaches, allows us to meaningfully talk about reducing (1) itself to yet another problem, or reducing another problem to it.

A relevant question here is whether and how the antecedent, as a resource, is allowed to be reused or "recycled". The answer to the *whether* part of this question is *yes* for both $\succ\!\!-$ and $\circ\!\!-$. Turing reduction does not impose any limits on how many times the oracle can be queried. Similarly, if our strategy for (1) had a need to repeatedly ask "*is H provable in predicate calculus?*" (for various $H$s), it would have been able to do so. This essentially means that multiple copies rather than a single copy of $A$ are played in the antecedent of $A\succ\!\!- B$ or $A\circ\!\!- B$.

As an aside, computability logic, of course, does have a stronger reduction operation which forbids repeated usage of the antecedent. The symbol for it is $\rightarrow$. In fact, $\rightarrow$ is considered the basic sort of reduction: $A\succ\!\!- B$ and $A\circ\!\!- B$ are defined through it as $\lambda A \rightarrow B$ and $\wp A \rightarrow B$, with $\lambda$ and $\wp$ being two flavors of what CL calls *recurrence* (=reusage) operations, both reminiscent of the *storage* operator ! of linear logic.[5] Not surprisingly, the logical behavior of $\rightarrow$ turns out to be different from that of $\succ\!\!-$ and $\circ\!\!-$ even in isolation: informally speaking, the former is (plainly) resource-sensitive while the latter are not. The following classical tautology

$$\bigl(P \supset (Q \supset R)\bigr) \supset \bigl((P \supset Q) \supset (P \supset R)\bigr)$$

is an example of an implicative formula which is valid with both $\succ\!\!-$ and $\circ\!\!-$ in the role of $\supset$ but invalid with $\rightarrow$. And an example of a tautology invalid with all three reduction operators $\supset\ \in\{\rightarrow,\succ\!\!-,\circ\!\!-\}$ is

$$\bigl((P \supset Q) \supset P\bigr) \supset P$$

which, known as *Pierce's Law*, is the shortest formula separating the implicative fragments of classical and intuitionistic logics.

Back to the resource reusage question, it is its *how* part where $\succ\!\!-$ and $\circ\!\!-$ start to differ from each other. $A\circ\!\!- B$ allows reusage of $A$ in the strongest possible sense, which makes $\circ\!\!-$ the weakest possible form of reduction, putting $\succ\!\!-$ strictly between $\circ\!\!-$ and $\rightarrow$. Both $A\succ\!\!- B$ and $A\circ\!\!- B$ — more precisely, their $\lambda A$ and $\wp A$ parts on which we are currently focused — can be characterized as games where the machine is allowed to restart the resource game $A$ an unlimited number of times without abandoning the already-in-progress sessions of $A$, thus forcing the environment to play/solve multiple parallel copies of $A$. The difference is that in $A\circ\!\!- B$, unlike $A\succ\!\!- B$, the machine does not have to restart $A$ from the very beginning every time it wants to reuse it; rather, the machine is (essentially) allowed to backtrack to any of the previous — not necessarily starting — positions and try a new continuation from there, thus depriving the adversary of the possibility to reconsider the moves it has already made in that position. This is in fact the type of reusage every purely software resource allows or would allow in the presence of an advanced operating system and unlimited memory: one can start running process $A$; then fork it at any stage thus creating two threads that have a common past but possibly diverging futures (with the possibility to treat one of the threads as a "backup copy" and preserve it for backtracking purposes); then further fork any of the branches at any time; and so on. The less flexible type of reusage of $A$ assumed by $A \succ\!\!- B$, on the other hand, is closer to what infinitely many autonomous physical resources would naturally offer, such as an unlimited number of independently acting robots each performing task $A$, or an unlimited number of computers with limited memories, each one only capable of and responsible for running a single thread of process $A$. Here the effect of replicating/forking an advanced stage of $A$ cannot be achieved unless, by good luck, there are two identical copies of the stage, meaning that the corresponding two robots or computers have so far acted in precisely the same ways.

---

[5]As the closest relative of our $\wp$ (but not $\lambda$) should be named Blass's [1, 2] semantically introduced *repetition operation* $R$ rather than Girard's syntactically introduced storage operator !. In very abstract, intuitive terms, $R$ could be characterized in a way rather similar to our characterization of $\wp$ given shortly. This means that the general game philosophy behind $\wp$ is largely the same as that behind $R$. However, when it comes to a materialization of that philosophy — to the technical level, that is — the similarity between the two operations is lost. The difference is partly related to the fact that $R$ is only applicable to a limited sort of games called *strict* in CL, while $\wp$ is applicable to everything we would potentially call an interactive computational task. Such a difference is somewhat reminiscent of the earlier-discussed difference in the degrees of generality between Turing reduction and our $\succ\!\!-$, $\circ\!\!-$, even though $R$, unlike Turing reduction, is on the interactive side, of course. This is by no means a historical survey paper, and therefore we refer to Sections 13 and 27 of [6] for a further discussion of how $\wp$ and $R$ compare. Note that the symbol used in [6] for $\wp$ was !.

As already pointed out, however, this difference between $A \succ\!\!- B$ and $A \circ\!\!- B$, while substantial in the general (truly interactive) case, is too subtle to be relevant when $A$ is a game that models only a very short and simple potential dialogue between the interacting parties, consisting in just asking a question and giving an answer. That is, when $A$ is from the class of problems to which the scopes of the ordinary concepts of computability, decidability or reducibility have been traditionally limited. And this is exactly why both of the two — $\succ\!\!-$ and $\circ\!\!-$ — versions of reducibility, when restricted to those special sorts of problems, fully coincide with the kind old textbook concept of Turing reducibility. As noted, the benefits from the greater degree of resource-reusage flexibility offered by $A \circ\!\!- B$ (as opposed to $A \succ\!\!- B$) are related to the possibility for the machine to try different reactions to the same action(s) by the environment in $A$. But such potential benefits cannot be realized when $A$ is, say, $\sqcap x \bigl(Pr(x) \sqcup \neg Pr(x)\bigr)$. Because here a given individual session of $A$ immediately ends with an environment's move, to which the machine simply has no legal or meaningful responses at all, let alone having multiple possible responses to experiment with.

Some readers might want to stop at this point and go no farther. That is because the rest of this paper is only meant for those who are already sufficiently familiar with computability logic, or are willing to do some parallel reading. CL has been introduced and reintroduced several times already ([6, 7, 8, 9, 10, 11]), and it is neither technically nor ethically feasible any more to do so once again, or even to reproduce all relevant formal definitions. Such definitions, without accompanying explanations and illustrations, may not be very helpful in getting a necessary degree of a feel of the subject, anyway. This article does rely, both technically and motivationally, on some prior material. Specifically, the proofs presented here should be read after or in parallel with the relevant parts of [10], which serves as a source of all special notation, terminology and concepts for the present paper (call it an appendix if you wish). While long, [10] is easy to read as it is written in a semitutorial style, without assuming any preliminary knowledge of the subject. Every unfamiliar term or notation used but not defined in the present paper can and should be looked up in [10], which has a convenient index of all terms and symbols. Familiarity with or parallel reading of [10] is a necessary and sufficient condition for understanding the rest of this paper.

## 2 Affine logic

Some of the proofs given in this paper will rely on the fact of the soundness of affine logic with respect to the CL semantics, established in Section 11 of [10]. Below we reproduce the relevant — propositional, unit-free, multiplicative-exponential — fragment of affine logic written in the symbols of CL. We follow the notational conventions of [10], according to which $\underline{G}$ stands for any (possible empty) finite sequence of formulas, $\mathord{\curlyvee}\underline{G}$ for any finite sequence of $\mathord{\curlyvee}$-prefixed formulas, etc. The nonunderlined letters $D, E, F, G, H, K$ stand just for formulas. *Sequents* in this version are one-sided, i.e. understood as just sequences of formulas rather than pairs of such sequences.

The **axioms** of (our present fragment of) affine logic are all sequents of the form

$$\neg E, E,$$

and the **rules of inference** are given by the following schemata:

**Exchange**

$$\frac{\underline{G}, E, F, \underline{H}}{\underline{G}, F, E, \underline{H}}$$

**Weakening**

$$\frac{\underline{G}}{\underline{G}, E}$$

**$\mathord{\curlyvee}$-Contraction**

$$\frac{\underline{G}, \mathord{\curlyvee} E, \mathord{\curlyvee} E}{\underline{G}, \mathord{\curlyvee} E}$$

**$\mathord{\curlywedge}$-Contraction**

$$\frac{\underline{G}, \mathord{\curlywedge} E, \mathord{\curlywedge} E}{\underline{G}, \mathord{\curlywedge} E}$$

**$\vee$-Introduction**

$$\frac{\underline{G}, E_1, \ldots, E_n}{\underline{G}, E_1 \vee \ldots \vee E_n}$$

**$\wedge$-Introduction**

$$\frac{\underline{G_1}, E_1 \quad \cdots \quad \underline{G_n}, E_n}{\underline{G_1}, \ldots, \underline{G_n}, E_1 \wedge \ldots \wedge E_n}$$

**$\mathord{\curlyvee}$-Introduction**

$$\frac{\underline{G}, E}{\underline{G}, \mathord{\curlyvee} E}$$

**$\mathord{\curlywedge}$-Introduction**

$$\frac{\underline{G}, E}{\underline{G}, \mathord{\curlywedge} E}$$

$$\text{\textwedge-Introduction} \qquad\qquad \text{\textvee-Introduction}$$

$$\frac{\curlyvee G, E}{\curlyvee G, \curlywedge E} \qquad\qquad \frac{\talloblong G, E}{\talloblong G, \between E}$$

This formulation does not officially forbid $\to$ in formulas, but it treats $E \to F$ as an abbreviation of $\neg E \vee F$, which explains why there are no rules for $\to$. More precisely, derivability of a formula containing $\to$ is to be understood as derivability of the result of eliminating $\to$ from that formula according to the above definition/prescription. Similarly, $E \succ\!\!- F$ is understood as $\neg {\curlywedge} E \vee F$ (i.e. ${\curlywedge} E \to F$) and $E \circ\!\!- F$ as $\neg {\between} E \vee F$ (i.e. ${\between} E \to F$). And, when applied to nonatomic formulas, $\neg$ is understood as an abbreviation defined by $\neg\neg E = E$, $\neg(E_1 \vee \ldots \vee E_n) = \neg E_1 \wedge \ldots \wedge \neg E_n$, $\neg(E_1 \wedge \ldots \wedge E_n) = \neg E_1 \vee \ldots \vee \neg E_n$, $\neg {\curlywedge} E = {\curlyvee} \neg E$, $\neg {\curlyvee} E = {\curlywedge} \neg E$, $\neg {\between} E = {\talloblong} \neg E$, $\neg {\talloblong} E = {\between} \neg E$.

Affine logic, in its full ($\bot, \top, \sqcup, \sqcap, \sqcap, \sqcup$-containing) language, was proven to be sound in [10] in the strong sense that, whenever a formula $E$ is provable in it, $E$ is not only valid but also uniformly valid, and that, furthermore, there is an effective procedure that takes an arbitrary proof of an arbitrary formula and constructs a uniform solution for that formula. That result, of course, automatically remains true for our present fragment of affine logic. This fragment is all we need in this paper, and henceforth we refer to it as *affine logic*, even though it is only a fragment of the latter in the proper sense.

The difference between our version of affine logic and Girard's [4] affine logic is minor, related to the presence of the two — ${\curlywedge}, {\curlyvee}$ and ${\between}, {\talloblong}$ — groups of recurrence ("exponential") operators rather than one group. Such a difference is not important because these groups of exponentials, having exactly the same inference rules, are deductively indistinguishable (from each other and from Girard's $!, ?$).

A known ([4]) result for affine logic straightforwardly applies to our version of it, according to which the logic is closed under the rule

$$\textbf{Cut}$$

$$\frac{\underline{G}, E \qquad \neg E, \underline{H}}{\underline{G}, \underline{H}}$$

Understanding $E \to F$, $E \succ\!\!- F$ and $E \circ\!\!- F$ as abbreviations of $\neg E \vee F$, $\neg {\curlywedge} E \vee F$ and $\neg {\between} E \vee F$ (but not forbidding $\neg$ to be applied to compound formulas), an occurrence $O$ of a subformula in a given formula, as usual, is said to be **negative** iff it is in the scope of an odd number of occurrences of $\neg$; otherwise it is **positive**. The following lemma is true no matter whether $\to, \succ\!\!-, \circ\!\!-$ are considered primitive symbols or not, or whether $\neg$ is allowed to be applied to compound formulas or not.

**Lemma 2.1** *Let $G_1$, $G_2$, $H_1$, $H_2$ be arbitrary formulas of the language of affine logic, such that the following two conditions are satisfied:*

- $\Vdash G_1 \to G_2$;

- $G_1$ *is a subformula of $H_1$, and $H_2$ is the result of replacing in $H_1$ a certain occurrence $O$ (fix it) of $G_1$ by $G_2$.*

*Then we have:*
  *a) If the occurrence $O$ is positive, then $\Vdash H_1 \to H_2$.*
  *b) If the occurrence $O$ is negative, then $\Vdash H_2 \to H_1$.*

**Proof.** Assume the conditions of the lemma (those that go before "Then we have:"). We also may and will assume that all of the connectives of $H_1$ are among $\neg, \downarrow\!\!\circ, \curlywedge$ and $\wedge$. Indeed, if this is not the case, then $H_1$ can be replaced by $H_1'$, where $H_1'$ is the result of eliminating the unwanted connectives trough rewriting each subformula $E_1 \vee \ldots \vee E_n$ as $\neg(\neg E_1 \wedge \ldots \wedge \neg E_n)$, each subformula $\circ\!\!\uparrow E$ as $\neg\!\downarrow\!\!\circ\neg E$, each subformula $E \to F$ as $\neg E \vee F$ (and then as $\neg(E \wedge \neg F)$), etc. Based on what we know from [10], for every interpretation $*$ we would have $H_1^* = H_1'^*$ (yes, equality rather than just equivalence). And the positive/negative status of any (occurrence of a) subformula of $H_1$ would remain the same in $H_1'$. To summarize, $H_1$ and $H_1'$ would be "the same" in every aspect relevant to our lemma, and hence could be safely identified in this proof.

We proceed by induction on the complexity of $H_1$ — more precisely, the complexity of $H_1$ minus the complexity of $G_1$, for the complexity of the $G_1$ part of $H_1$ is irrelevant. In this proof we will explicitly or implicitly rely on the fact of the closure of uniform validity under modus ponens proven in Section 13 of [10], and the already mentioned fact of the soundness of affine logic with respect to uniform validity.

If $H_1 = G_1$, then $H_2 = G_2$. Here the occurrence of $G_1$ in $H_1$ (i.e. in itself) is positive, and we have $\Vdash H_1 \to H_2$ just by our assumption that $\Vdash G_1 \to G_2$.

Next, assume $H_1 = \neg E_1$, and the occurrence $O$ of $G_1$ in $H_1$ is the occurrence $O'$ of $G_1$ in $E_1$. Let $E_2$ be the result of replacing in $E_1$ the occurrence $O'$ by $G_2$. Of course, $H_2 = \neg E_2$. Suppose $O$ is positive. Then $O'$ is negative. By the induction hypothesis, $\Vdash E_2 \to E_1$. It is an easy syntactic exercise to verify that affine logic proves $(E_2 \to E_1) \to (\neg E_1 \to \neg E_2)$, i.e. $(E_2 \to E_1) \to (H_1 \to H_2)$. Hence $\Vdash (E_2 \to E_1) \to (H_1 \to H_2)$ and, by (the closure of uniform validity under) modus ponens, $\Vdash H_1 \to H_2$. The case when $O$ is negative will be handled in a symmetric way.

Next, assume $H_1 = \downarrow\!\!\circ E_1$, and the occurrence $O$ of $G_1$ in $H_1$ is the occurrence $O'$ of $G_1$ in $E_1$. Let $E_2$ be the result of replacing in $E_1$ the occurrence $O'$ by $G_2$. We thus have $H_2 = \downarrow\!\!\circ E_2$. Suppose $O$ is positive. Then so is $O'$. By the induction hypothesis, $\Vdash E_1 \to E_2$. Then, according the $\downarrow\!\!\circ$-closure lemma proven in Section 13 of [10], $\Vdash \downarrow\!\!\circ(E_1 \to E_2)$. One can routinely verify that affine logic derives $\downarrow\!\!\circ(E_1 \to E_2) \to (\downarrow\!\!\circ E_1 \to \downarrow\!\!\circ E_2)$, i.e. $\downarrow\!\!\circ(E_2 \to E_1) \to (H_1 \to H_2)$. Hence $\Vdash \downarrow\!\!\circ(E_2 \to E_1) \to (H_1 \to H_2)$ and, by modus ponens, $\Vdash H_1 \to H_2$. The case when $O$ is negative will be handled in a symmetric way.

The case $H_1 = \curlywedge E_1$ is similar to the previous one, with the only difference that it relies on the $\curlywedge$-closure lemma rather than the $\downarrow\!\!\circ$-closure lemma of Section 13 of [10].

Finally, assume that ($H_1 \neq G_1$ and) the main operator of $H_1$ is $\wedge$. As an example, here we only consider the case $H_1 = E_1 \wedge F$, where the occurrence $O$ of $G_1$ in $H_1$ is the occurrence $O'$ of $G_1$ in $E_1$. All other cases will be similar. So, we have $H_2 = E_2 \wedge F$, where $E_2$ is the result of replacing in $E_1$ the occurrence $O'$ by $G_2$. Suppose $O$ is positive. Then so is $O'$. By the induction hypothesis, $\Vdash E_1 \to E_2$. One can also verify that affine logic proves $(E_1 \to E_2) \to (E_1 \wedge F \to E_2 \wedge F)$, i.e. $(E_1 \to E_2) \to (H_1 \to H_2)$. Hence $\Vdash (E_1 \to E_2) \to (H_1 \to H_2)$ and, by modus ponens, $\Vdash H_1 \to H_2$. The case when $O$ is negative will be handled in a symmetric way. □

## 3 Intuitionistic logic

### 3.1 Syntax

The implicative fragment of intuitionistic logic comes to us in two versions: $\mathbf{Int}^{\circ\!-}$, which reads intuitionistic implication as $\circ\!-$, and $\mathbf{Int}^{\succ\!-}$, which reads it as $\succ\!-$. Its language has infinitely many propositional (i.e. 0-ary) letters (=atoms), for which we use the metavariables $P$, $Q$, $R$, $X$, $Y$, $Z$, $W$, possibly with indices. It should be noted that, when it comes to interpretations, these are *general* rather than elementary letters (see [10], Section 7), just as the letters of affine logic are. The formulas of the language of $\mathbf{Int}^{\circ\!-}$, to which we refer as $\mathbf{Int}^{\circ\!-}$**-formulas**, are built from atoms in the standard way using the binary operator $\circ\!-$. There are no other operators in the language, nor are there any logical atoms such as $\top$ or $\bot$. Similarly for the language of $\mathbf{Int}^{\succ\!-}$, with $\mathbf{Int}^{\succ\!-}$**-formulas** using $\succ\!-$ instead of $\circ\!-$. The languages of $\mathbf{Int}^{\circ\!-}$ and $\mathbf{Int}^{\succ\!-}$ are thus sublanguages of the language of affine logic.

Next, an $\mathbf{Int}^{\circ\!-}$**-sequent** is a pair $\underline{G} \Rightarrow E$, where $\underline{G}$, called the **antecedent**, is any finite sequence of $\mathbf{Int}^{\circ\!-}$-formulas, and $E$, called the **succedent**, is a (one single) $\mathbf{Int}^{\circ\!-}$-formula. $\mathbf{Int}^{\succ\!-}$**-sequents** are defined similarly. Unlike our choice of one-sided sequents for affine logic, here we thus deal with two-sided sequents.

Below is a Gentzen-style deductive system for $\mathbf{Int}^{\circ\!-}$. A formula $K$ is considered provable in it iff the empty-antecedent sequent $\Rightarrow K$ is provable.

The **axioms** of $\mathbf{Int}^{\circ\!-}$ are all $\mathbf{Int}^{\circ\!-}$-sequents of the form $K \Rightarrow K$, and the **rules of inference** are:

| Exchange | Weakening | Contraction |
|---|---|---|
| $\dfrac{\underline{G}, E, F, \underline{H} \Rightarrow K}{\underline{G}, F, E, \underline{H} \Rightarrow K}$ | $\dfrac{\underline{G} \Rightarrow K}{\underline{G}, E \Rightarrow K}$ | $\dfrac{\underline{G}, F, F \Rightarrow K}{\underline{G}, F \Rightarrow K}$ |

| **Right** ∘— | **Left** ∘— |
|:---:|:---:|
| $\underline{G}, F \Rightarrow K$ | $\underline{G}, F \Rightarrow K_1 \qquad \underline{H} \Rightarrow K_2$ |
| $\underline{G} \Rightarrow F \mathbin{\circ\mkern-3mu-} K$ | $\underline{G}, \underline{H}, K_2 \mathbin{\circ\mkern-3mu-} F \Rightarrow K_1$ |

The system $\mathbf{Int}^{\succ}$ is defined in the same way as $\mathbf{Int}^{\circ\mkern-3mu-}$, only with $\succ$ instead of $\circ\mkern-3mu-$, of course.

## 3.2 Kripke semantics

A **Kripke model** [13] is a triple $\mathcal{M} = (\mathcal{W}, \mathcal{R}, \models)$, where:

- $\mathcal{W}$ is a (here) finite set of what are called **worlds** (of $\mathcal{M}$).

- $\mathcal{R}$ is a transitive and reflexive relation between worlds. When $p\mathcal{R}q$, we say that world $q$ is **accessible** (in $\mathcal{M}$) from world $p$.

- $\models$ is a relation between worlds and $\mathbf{Int}^{\circ\mkern-3mu-}$-formulas, satisfying the following two conditions for all formulas $E, F$ and worlds $p, q$:

    - if $p \models E$ and $p\mathcal{R}q$, then $q \models E$;
    - $p \models E \mathbin{\circ\mkern-3mu-} F$ iff, whenevr $p\mathcal{R}q$ and $q \models E$, we have $q \models F$.

    When $\underline{G}$ is a sequence or a set of formulas, we write $p \models \underline{G}$ to mean that $p \models G$ for every formula $G$ of $\underline{G}$. The relation $\models$ further extends to sequents by stipulating that $p \models \underline{G} \Rightarrow E$ iff, for every world $q$ accessible from $p$, if $q \models \underline{G}$, then $q \models E$. The symbol $\not\models$, as expected, will be used for the negation of $\models$.

Let $\mathcal{M} = (\mathcal{W}, \mathcal{R}, \models)$ be a Kripke model. Where $\mathcal{F}$ is a $\mathbf{Int}^{\circ\mkern-3mu-}$-formula, or a sequence of $\mathbf{Int}^{\circ\mkern-3mu-}$-formulas, or a set of $\mathbf{Int}^{\circ\mkern-3mu-}$-formulas, or a $\mathbf{Int}^{\circ\mkern-3mu-}$-sequent, we write $\mathcal{M} \models \mathcal{F}$ to mean that $p \models \mathcal{F}$ for every $p \in \mathcal{W}$. And we say that an $\mathbf{Int}^{\circ\mkern-3mu-}$-formula $E$ is $\mathcal{M}$-**equivalent** to an $\mathbf{Int}^{\circ\mkern-3mu-}$-formula $F$ — symbolically $\mathcal{M} \models E \circ\!\!-\!\!\circ F$ — iff $\mathcal{M} \models E \mathbin{\circ\mkern-3mu-} F$ and $\mathcal{M} \models F \mathbin{\circ\mkern-3mu-} E$. Obviously $\mathcal{M} \models E \circ\!\!-\!\!\circ F$ means that, for each world $p$ of $\mathcal{M}$, $p \models E$ iff $p \models F$, so that "$\mathcal{M}$ does not see any difference between $E$ and $F$". Note also that whenever $\mathcal{M} \models E \circ\!\!-\!\!\circ E'$ and $\mathcal{M} \models F \circ\!\!-\!\!\circ F'$, we also have $\mathcal{M} \models (E \mathbin{\circ\mkern-3mu-} F) \circ\!\!-\!\!\circ (E' \mathbin{\circ\mkern-3mu-} F')$.

We read $\xi \models \varsigma$ (whatever $\xi$ and $\varsigma$ are) as "$\varsigma$ is **true** in $\xi$", and $\xi \not\models \varsigma$ as "$\varsigma$ is **false** in $\xi$".

It is an established ([13]) fact that $\mathbf{Int}^{\circ\mkern-3mu-}$ is sound and complete with respect to Kripke semantics, in the sense that an $\mathbf{Int}^{\circ\mkern-3mu-}$-sequent or formula $\mathcal{S}$ is provable in $\mathbf{Int}^{\circ\mkern-3mu-}$ if (**completeness**) and only if (**soundness**) $\mathcal{S}$ is true in every Kripke model. For known and straightforward reasons, Kripke models here can be restricted to ones where $(\mathcal{W}, \mathcal{R})$ forms a tree rather than just a partial order, and the above fact then can be rephrased by saying that a sequent or formula is provable in $\mathbf{Int}^{\circ\mkern-3mu-}$ iff it is true in the root world of every tree-like Kripke model.

## 3.3 Main theorem

By a **$\Sigma_1$-predicate** we mean a predicate that can be written as $\exists zA$ for some decidable finitary predicate (elementary game) $A$. And a **Boolean combination** of $\Sigma_1$-predicates is a $\neg, \wedge, \vee$-combination of such games.

Note that, since we deal with a propositional language, every interpretation is admissible for every formula. So, when considering interpretations, there is no need to bother about or even mention their admissibility.

Below is our main theorem, according to which the implicative fragment of intuitionistic logic is sound and complete with respect to the semantics of computability logic, no matter whether the intuitionistic implication is read as $\succ\!\!-$ or $\circ\!\!-$. Both the soundness and the completeness clauses come in certain strong forms.

**Theorem 3.1** *For any $\mathbf{Int}^{\circ\!-}$- (resp. $\mathbf{Int}^{\succ\!-}$-) formula $K$, $K$ is provable in $\mathbf{Int}^{\circ\!-}$ (resp. $\mathbf{Int}^{\succ\!-}$) iff $K$ is valid iff $K$ is uniformly valid. Furthermore:*

*(a) There is an effective procedure that takes an arbitrary $\mathbf{Int}^{\circ\!-}$- (resp. $\mathbf{Int}^{\succ\!-}$-) proof of an arbitrary formula $K$ and constructs a uniform solution for $K$.*

*(b) If an $\mathbf{Int}^{\circ\!-}$- (resp. $\mathbf{Int}^{\succ\!-}$-) formula $K$ is not provable in $\mathbf{Int}^{\circ\!-}$ (resp. $\mathbf{Int}^{\succ\!-}$), then $K^*$ is not computable for some interpretation $*$ satisfying the following **complexity condition**: $*$ interprets every atom (of the languages of $\mathbf{Int}^{\circ\!-}$, $\mathbf{Int}^{\succ\!-}$) as a problem of the form $\sqcup xB$, where $B$ is a Boolean combination of $\Sigma_1$-predicates.*

**Proof.** As already mentioned and relied upon, affine logic is sound in the strong sense of clause (a) of this theorem ([10], Section 11). So, to prove the soundness of $\mathbf{Int}^{\circ\!-}$, $\mathbf{Int}^{\succ\!-}$ and the corresponding clause (a), it would be sufficient to show that affine logic derives any formula $K$ that $\mathbf{Int}^{\circ\!-}$ or $\mathbf{Int}^{\succ\!-}$ does, and that, furthermore, that an affine-logic proof of $K$ can be effectively constructed from an $\mathbf{Int}^{\circ\!-}$- or $\mathbf{Int}^{\succ\!-}$-proof of it. This purely syntactic and easy-to-verify (by induction on the lengths of proofs) fact can be considered already known in the form of Girard's [4] embedding of intuitionistic calculus into linear (and hence also affine) logic. Indeed, unlike intuitionistic disjunction for which Girard's reading is essentially different from ours,[6] intuitionistic implication from $F$ to $G$ in [4] is understood as $!F \rightarrow G$, just like we read it as $\flat F \rightarrow G$ or $\lambda F \rightarrow G$. Of course, the soundness of $\mathbf{Int}^{\circ\!-}$ also immediately follows from the soundness of the full intuitionistic calculus proven in [11].

As for the completeness part of Theorem 3.1 and the corresponding clause (b), this is exactly to what the rest of the present paper is devoted. □

# 4 Machines playing against machines

This section borrows a discussion from [7], providing certain background information necessary for our completeness proof but missing in [10], the only external source on computability logic on which the present paper was promised to rely.

---

[6]Girard's translation for the intuitionistic disjunction of $F$ and $G$ is $!F \oplus !G$, while our reading is (simply) $F \sqcup G$. This explains why one could not apply a similar soundness argument to $\mathbf{Int}^{\succ\!-}$ in the full (disjunction-containing) language, which, as mentioned in Section 1, can indeed be shown to be unsound.

Remember that $\neg\Gamma$, when $\Gamma$ is a run, means the result of reversing all labels in $\Gamma$. For a run $\Gamma$ and a computation branch $B$ of an HPM or EPM, we say that $B$ **cospells** $\Gamma$ iff $B$ spells $\neg\Gamma$ in the sense of Section 6 of [10]. Intuitively, when a machine $\mathcal{M}$ plays as $\bot$ (rather than $\top$), then the run that is generated by a given computation branch $B$ of $\mathcal{M}$ is the run cospelled (rather than spelled) by $B$, for the moves that $\mathcal{M}$ makes get the label $\bot$, and the moves that its adversary makes get the label $\top$.

We say that an EPM $\mathcal{E}$ is **fair** iff, for every valuation $e$, every $e$-computation branch of $\mathcal{E}$ is fair in the sense of Section 6 of [10].

**Lemma 4.1** *Assume $\mathcal{E}$ is a fair EPM, $\mathcal{H}$ is any HPM, and $e$ is any valuation. There are a uniquely defined $e$-computation branch $B_\mathcal{E}$ of $\mathcal{E}$ and a uniquely defined $e$-computation branch $B_\mathcal{H}$ of $\mathcal{H}$ — which we respectively call **the $(\mathcal{E}, e, \mathcal{H})$-branch** and **the $(\mathcal{H}, e, \mathcal{E})$-branch** — such that the run spelled by $B_\mathcal{H}$, called **the $\mathcal{H}$ vs. $\mathcal{E}$ run on** $e$, is the run cospelled by $B_\mathcal{E}$.*

When $\mathcal{H}, \mathcal{E}, e$ are as above, $\Gamma$ is the $\mathcal{H}$ vs. $\mathcal{E}$ run on $e$ and $A$ is a game with $\mathbf{Wn}_e^A\langle\Gamma\rangle = \top$ (resp. $\mathbf{Wn}_e^A\langle\Gamma\rangle = \bot$), we say that $\mathcal{H}$ **wins** (resp. **loses**) $A$ **against** $\mathcal{E}$ **on** $e$.

A strict proof of the above lemma can be found in [6] (Lemma 20.4), and we will not reproduce the formal proof here. Instead, the following intuitive explanation would suffice:

**Proof idea.** Assume $\mathcal{H}, \mathcal{E}, e$ are as in Lemma 4.1. The play that we are going to describe is the unique play generated when the two machines play against each other, with $\mathcal{H}$ in the role of $\top$, $\mathcal{E}$ in the role of $\bot$, and $e$ spelled on the valuation tapes of both machines. We can visualize this play as follows. Most of the time during the process $\mathcal{H}$ remains inactive (sleeping); it is woken up only when $\mathcal{E}$ enters a permission state, on which event $\mathcal{H}$ makes a (one single) transition to its next computation step — that may or may not result in making a move — and goes back to sleep that will continue until $\mathcal{E}$ enters a permission state again, and so on. From $\mathcal{E}$'s perspective, $\mathcal{H}$ acts as a patient adversary who makes one or zero move only when granted permission, just as the EPM-model assumes. And from $\mathcal{H}$'s perspective, who, like a person in a comma, has no sense of time during its sleep and hence can think that the wake-up events that it calls the beginning of a clock cycle happen at a constant rate, $\mathcal{E}$ acts as an adversary who can make any finite number of moves during a clock cycle (i.e. while $\mathcal{H}$ was sleeping), just as the HPM-model assumes. This scenario uniquely determines an $e$-computation branch $B_\mathcal{E}$ of $\mathcal{E}$ that we call the $(\mathcal{E}, e, \mathcal{H})$-branch, and an $e$-computation branch $B_\mathcal{H}$ of $\mathcal{H}$ that we call the $(\mathcal{H}, e, \mathcal{E})$-branch. What we call the $\mathcal{H}$ vs. $\mathcal{E}$ run on $e$ is the run generated in this play. In particular — since we let $\mathcal{H}$ play in the role of $\top$ — this is the run spelled by $B_\mathcal{H}$. $\mathcal{E}$, who plays in the role of $\bot$, sees the same run, only it sees the labels of the moves of that run in negative colors. That is, $B_\mathcal{E}$ cospells rather than spells that run. This is exactly what Lemma 4.1 asserts. $\diamond$

# 5 Standardization

Let us agree that, throughout this section, "formula" means $\mathbf{Int}^{\circ\!\!-}$-formula, "sequent" means $\mathbf{Int}^{\circ\!\!-}$-sequent, and "provable" ($\vdash$) means provable in $\mathbf{Int}^{\circ\!\!-}$.

We say that a sequent is **standard** iff it is

$$\left.\begin{array}{l} X_1 \circ\!\!-(Y_1 \circ\!\!- Z_1),\ \cdots,\ X_s \circ\!\!-(Y_s \circ\!\!- Z_s), \\ (P_1 \circ\!\!- Q_1) \circ\!\!- R_1,\ \cdots,\ (P_s \circ\!\!- Q_s) \circ\!\!- R_s \end{array}\right\} \Rightarrow W,$$

where $s \geq 0$ and the $X_i, Y_i, Z_i, P_i, Q_i, R_i$ ($1 \leq i \leq s$) and $W$ are atoms.

Where $K$ is a formula and $H$ is a subformula of it, throughout this section we will be using the notation $H^K$ for a certain atom which, intuitively, is a "standard atomic name" assigned by us to $H$. Specifically, let $G_1, \ldots, G_s$ be all of the non-atomic subformulas of $K$ listed according to the lexicographic order, and let $W_1, \ldots, W_s$ be the first (in the lexicographic order) $s$ atoms of the language of $\mathbf{Int}^{\circ\!\!-}$ not occurring in $K$. Then we define the atom $H^K$ by stipulating that:

- $H^K = H$ if $H$ is atomic;

- $H^K = W_i$ if $H = G_i$ ($1 \leq i \leq s$).

Let $K$ be a formula, and

$$G_1 = E_1 \circ\!\!- F_1,\ \ldots,\ G_s = E_s \circ\!\!- F_s$$

all of its non-atomic subformulas, listed according to the lexicographic order. We define the **standardization** of $K$ as the following sequent:

$$\left.\begin{array}{l} G_1^K \circ\!\!-(E_1^K \circ\!\!- F_1^K),\ \cdots,\ G_s^K \circ\!\!-(E_s^K \circ\!\!- F_s^K), \\ (E_1^K \circ\!\!- F_1^K) \circ\!\!- G_1^K,\ \cdots,\ (E_s^K \circ\!\!- F_s^K) \circ\!\!- G_s^K \end{array}\right\} \Rightarrow K^K.$$

Notice that the standardization of $K$ is (indeed) a standard sequent.

*Example*: Where $W_1, W_2, W_3$ are lexicographically the first three atoms different from $Q$ and $R$, the following sequent is the standardization of $Q \circ\!\!- ((Q \circ\!\!- R) \circ\!\!- R)$:

$$\left.\begin{array}{l} W_1 \circ\!\!-(Q \circ\!\!- R),\ W_2 \circ\!\!-(W_1 \circ\!\!- R),\ W_3 \circ\!\!-(Q \circ\!\!- W_2), \\ (Q \circ\!\!- R) \circ\!\!- W_1,\ (W_1 \circ\!\!- R) \circ\!\!- W_2,\ (Q \circ\!\!- W_2) \circ\!\!- W_3 \end{array}\right\} \Rightarrow W_3.$$

**Lemma 5.1** *Let $K$ be an arbitrary formula, $\underline{G} \Rightarrow W$ its standardization, and $\mathcal{M} = (\mathcal{W}, \mathcal{R}, \models)$ a Kripke model with $\mathcal{M} \models \underline{G}$. Then every subformula $H$ of $K$ is $\mathcal{M}$-equivalent to $H^K$. Consequently, $\mathcal{M} \models K \circ\!\!-\!\!\circ W$.*

**Proof.** Let $K$, $\underline{G} \Rightarrow W$, $\mathcal{M}$, $H$ be as the lemma assumes. We proceed by induction on the complexity of $H$. The case when $H$ is atomic is trivial, because then $H^K = H$. Suppose now $H = E \circ\!\!- F$. Then $\underline{G}$ contains $H^K \circ\!\!-(E^K \circ\!\!- F^K)$ and $(E^K \circ\!\!- F^K) \circ\!\!- H^K$. Therefore, as $\mathcal{M} \models \underline{G}$, we have

$$\mathcal{M} \models (E^K \circ\!\!- F^K) \circ\!\!-\!\!\circ H^K.$$

But, by the induction hypothesis, $\mathcal{M}\models E^K \circ\!\!-\!\!\circ E$ and $\mathcal{M}\models F^K \circ\!\!-\!\!\circ F$. Hence

$$\mathcal{M}\models (E^K \circ\!\!-\!\! F^K) \circ\!\!-\!\!\circ (E \circ\!\!-\!\! F).$$

Consequently, $\mathcal{M}\models (E \circ\!\!-\!\! F) \circ\!\!-\!\!\circ H^K$, i.e. $\mathcal{M}\models H \circ\!\!-\!\!\circ H^K$. □

**Lemma 5.2** *If [7] a formula is not provable, then its standardization is not provable, either.*

**Proof.** Consider an arbitrary formula $K$ and its standardization $\underline{G} \Rightarrow W$. Assume $\not\vdash K$. Then, by the completeness of $\mathbf{Int}^{\circ\!-}$ with respect to Kripke semantics, there is a model $\mathcal{M} = (\mathcal{W}, \mathcal{R}, \models)$ with $\mathcal{M} \not\models K$. We may assume here that, for every subformula $H$ of $K$, $\mathcal{M}\models H \circ\!\!-\!\!\circ H^K$. Indeed, when $H$ is atomic, then it is automatically $\mathcal{M}$-equivalent to $H^K$ because $H = H^K$. And if $H$ is not atomic, then $H^K$ is not among the atoms of $K$, and we may make arbitrary assumptions regarding in what worlds the atom $H^K$ is true without affecting the fact that $\mathcal{M} \not\models K$; so, our assumption is that $H^K$ is true exactly in the the worlds where $H$ is true.

We claim that $\mathcal{M}\models \underline{G}$. Indeed, pick an arbitrary formula $H$ of $\underline{G}$. We need to show that $\mathcal{M}\models H$. There are two cases to consider, depending on the form of $H$. One possibility is that $H = G^K \circ\!\!-\!\!(E^K \circ\!\!-\!\! F^K)$. By the above assumptions regarding $\mathcal{M}$, we have $\mathcal{M}\models G^K \circ\!\!-\!\!\circ G$, $\mathcal{M}\models E^K \circ\!\!-\!\!\circ E$ and $\mathcal{M}\models F^K \circ\!\!-\!\!\circ F$. So, in order to verify that $\mathcal{M}\models H$, it would suffice to show that $\mathcal{M}\models G \circ\!\!-\!\!(E \circ\!\!-\!\! F)$. But this is trivially so because — note — $G$ is nothing but $E \circ\!\!-\!\! F$. The other possibility is $H = (E^K \circ\!\!-\!\! F^K) \circ\!\!-\!\! G^K$, which is similar.

Now, as $\mathcal{M} \not\models K$, we have $\mathcal{M} \not\models W$ because, by Lemma 5.1, $\mathcal{M}\models W \circ\!\!-\!\!\circ K$. This, together with $\mathcal{M}\models \underline{G}$, means that $\mathcal{M}\not\models \underline{G} \Rightarrow W$. Therefore, by the soundness of $\mathbf{Int}^{\circ\!-}$ with respect to Kripke semantics, $\not\vdash \underline{G} \Rightarrow W$, as desired. □

**Lemma 5.3** *Let $K$ be an arbitrary formula, and $\underline{G} \Rightarrow W$ its standardization. Then $\vdash K, \underline{G} \Rightarrow W$.*

**Proof.** Assume, for a contradiction, that $\underline{G} \Rightarrow W$ is the standardization of a formula $K$, and $\not\vdash K, \underline{G} \Rightarrow W$. Then, by the completeness of $\mathbf{Int}^{\circ\!-}$ with respect to Kripke semantics, there is a Kripke model $\mathcal{M} = (\mathcal{W}, \mathcal{R}, \models)$ with $\mathcal{M} \not\models K, \underline{G} \Rightarrow W$, meaning that, for some world $p \in \mathcal{W}$, we have $p \models K$, $p \models \underline{G}$ and $p \not\models W$. Obviously here we may assume that every world is accessible from $p$, so that $\mathcal{M}\models \underline{G}$. Then, by Lemma 5.1, $\mathcal{M}\models K \circ\!\!-\!\!\circ W$. But this is a contradiction, because $p \models K$ and $p \not\models W$. □

## 6 Desequentization

In this section, we will write $\mathbf{Int}^{\circ\!-}\vdash$ for provability in $\mathbf{Int}^{\circ\!-}$ and $\mathbf{AL}\vdash$ for provability in affine logic.

---

[7] In fact, this lemma can be shown to be true in the stronger "if and only if" form, but for our purposes this is not necessary.

Let $\mathcal{S}$ be the (arbitrary) standard sequent

$$\left.\begin{array}{l} X_1 \circ\!\!-\, (Y_1 \circ\!\!-\, Z_1), \;\cdots,\; X_s \circ\!\!-\, (Y_s \circ\!\!-\, Z_s), \\ (P_1 \circ\!\!-\, Q_1) \circ\!\!-\, R_1, \;\cdots,\; (P_s \circ\!\!-\, Q_s) \circ\!\!-\, R_s \end{array}\right\} \;\Rightarrow\; W.$$

Then we define the **desequentization** of $\mathcal{S}$ as the following formula of the language of affine logic:

$$\begin{array}{l} \downarrow\!\!\circ\!(X_1 \wedge Y_1 \to Z_1) \wedge \;\cdots\; \wedge \downarrow\!\!\circ\!(X_s \wedge Y_s \to Z_s) \;\wedge \\ \downarrow\!\!\circ\!\bigl((\downarrow\!\!\circ P_1 \to Q_1) \to R_1\bigr) \wedge \;\cdots\; \wedge \downarrow\!\!\circ\!\bigl((\downarrow\!\!\circ P_s \to Q_s) \to R_s\bigr) \end{array} \;\to\; W.$$

As an aside, the desequentization of $\mathcal{S}$ should not be understood as expressing the "intended meaning" (say, as defined in [11]) of $\mathcal{S}$. A formula expressing the intended meaning of $\mathcal{S}$ can be obtained from the above one by inserting a $\downarrow\!\!\circ$ before every subformula $X_i$, $Y_i$ and $(\downarrow\!\!\circ P_i \to Q_i)$. $\mathcal{S}$ and its desequentization are *not* semantically equivalent: the latter, with its missing occurrences of $\downarrow\!\!\circ$, is weaker than the former.

**Lemma 6.1** *Assume $K$ is an $\mathbf{Int}^{\circ\!\!-}$-formula, and $D$ is the desequentization of the standardization of $K$. Then, for any interpretation $^*$ with $\models K^*$, we have $\models D^*$.*

**Proof.** Consider an arbitrary $\mathbf{Int}^{\circ\!\!-}$-formula $K$ and an arbitrary interpretation $^*$ with $\models K^*$. Let

$$\left.\begin{array}{l} X_1 \circ\!\!-\, (Y_1 \circ\!\!-\, Z_1), \;\cdots,\; X_s \circ\!\!-\, (Y_s \circ\!\!-\, Z_s), \\ (P_1 \circ\!\!-\, Q_1) \circ\!\!-\, R_1, \;\cdots,\; (P_s \circ\!\!-\, Q_s) \circ\!\!-\, R_s \end{array}\right\} \;\Rightarrow\; W \qquad (2)$$

be the standardization of $K$. Using $G_1, \ldots, G_{2s}$ as abbreviations of the $2s$ formulas of the antecedent of the above sequent, we rewrite it as

$$G_1, \ldots, G_{2s} \Rightarrow W.$$

According to Lemma 5.3,

$$\mathbf{Int}^{\circ\!\!-} \vdash K, G_1, \ldots, G_{2s} \Rightarrow W.$$

From here, applying the Right $\circ\!\!-$ rule $2s+1$ times, we get

$$\mathbf{Int}^{\circ\!\!-} \vdash K \circ\!\!-\, (G_1 \circ\!\!-\, (G_2 \circ\!\!-\, \ldots (G_{2s} \circ\!\!-\, W)\ldots)).$$

As noted in our proof of the soundness part of Theorem 3.1, affine logic proves every formula provable in $\mathbf{Int}^{\circ\!\!-}$. So, the above $\mathbf{Int}^{\circ\!\!-}$-provability translates into

$$\mathbf{AL} \vdash K \circ\!\!-\, (G_1 \circ\!\!-\, (G_2 \circ\!\!-\, \ldots (G_{2s} \circ\!\!-\, W)\ldots)). \qquad (3)$$

Next, in a routine syntactic exercise, one can show that

$$\mathbf{AL} \vdash \neg(K \circ\!\!-\, (G_1 \circ\!\!-\, (G_2 \circ\!\!-\, \ldots (G_{2s} \circ\!\!-\, W)\ldots))),\; \downarrow\!\!\circ K \to (\downarrow\!\!\circ G_1 \wedge \ldots \wedge \downarrow\!\!\circ G_{2s} \to W).$$

The above, together with (3), by the closure of affine logic under cut, implies

$$\mathbf{AL} \vdash {\downarrow}K \to ({\downarrow}G_1 \wedge \ldots \wedge {\downarrow}G_{2s} \to W),$$

whence, by the soundness of affine logic,

$$\models {\downarrow}K^* \to ({\downarrow}G_1 \wedge \ldots \wedge {\downarrow}G_{2s} \to W)^*. \tag{4}$$

According to the theorem of Section 10 of [10], the rule of modus ponens preserves computability. And, according to one of the lemmas of Section 13 of [10], so does the rule "*from $A$ to ${\downarrow}A$*". So, our assumption $\models K^*$ implies $\models {\downarrow}K^*$, and the latter, together with (4), by the closure of computability under modus ponens, implies

$$\models ({\downarrow}G_1 \wedge \ldots \wedge {\downarrow}G_{2s} \to W)^*.$$

If we now disabbreviate the $G_i$s, the formula ${\downarrow}G_1 \wedge \ldots \wedge {\downarrow}G_{2s} \to W$ rewrites as

$$\begin{array}{l} {\downarrow}\bigl({\downarrow}X_1 \to ({\downarrow}Y_1 \to Z_1)\bigr) \wedge \;\cdots\; \wedge {\downarrow}\bigl({\downarrow}X_s \to ({\downarrow}Y_s \to Z_s)\bigr) \;\wedge \\ {\downarrow}\bigl({\downarrow}({\downarrow}P_1 \to Q_1) \to R_1\bigr) \wedge \;\cdots\; \wedge {\downarrow}\bigl({\downarrow}({\downarrow}P_s \to Q_s) \to R_s\bigr) \end{array} \to W, \tag{5}$$

so we have $\models (5)^*$. Next, consider the formula

$$\begin{array}{l} {\downarrow}({\downarrow}X_1 \wedge {\downarrow}Y_1 \to Z_1) \wedge \;\cdots\; \wedge {\downarrow}({\downarrow}X_s \wedge {\downarrow}Y_s \to Z_s) \;\wedge \\ {\downarrow}\bigl({\downarrow}({\downarrow}P_1 \to Q_1) \to R_1\bigr) \wedge \;\cdots\; \wedge {\downarrow}\bigl({\downarrow}({\downarrow}P_s \to Q_s) \to R_s\bigr) \end{array} \to W. \tag{6}$$

One can easily verify that affine logic proves $(5) \to (6)$ and hence, by the soundness of affine logic, $\models (5)^* \to (6)^*$. From here and $\models (5)^*$, by the fact that modus ponens preserves computability, we infer

$$\models (6)^*.$$

Next, for any formula $E$, of course, $\mathbf{AL} \vdash {\downarrow}E \to E$. Hence, by the soundness of affine logic,

$$\Vdash {\downarrow}E \to E \quad (any\ E). \tag{7}$$

Now, consider the desequentization $D$ of (2), which is

$$\begin{array}{l} {\downarrow}(X_1 \wedge Y_1 \to Z_1) \wedge \;\cdots\; \wedge {\downarrow}(X_s \wedge Y_s \to Z_s) \;\wedge \\ {\downarrow}\bigl(({\downarrow}P_1 \to Q_1) \to R_1\bigr) \wedge \;\cdots\; \wedge {\downarrow}\bigl(({\downarrow}P_s \circ\!\!-\, Q_s) \to R_s\bigr) \end{array} \to W.$$

Observe that $D$ is nothing but the result of deleting ${\downarrow}$ before every subformula $X_i$, $Y_i$ and $({\downarrow}P_i \to Q_i)$ in (6). And (the occurrences of) all such subformulas are positive. So, with (7) in mind, we can rely — $3s$ times — on Lemma 2.1 in combination with the Transitivity lemma of Section 13 of [10], and conclude that $\Vdash (6) \to D$ [8] and hence $\models (6)^* \to D^*$. This, together with the earlier established fact $\models (6)^*$, by the closure of computability under modus ponens, implies the desired $\models D^*$. □

---

[8] Of course, we could have shown the same by verifying that (the sound) affine logic proves $(6) \to D$.

# 7  Main lemma

**Lemma 7.1** *If a standard $\mathbf{Int}^{\circ\!-}$-sequent is not provable in $\mathbf{Int}^{\circ\!-}$, then its desequentization $D$ is not valid; specifically, there is an interpretation $*$ satisfying the complexity condition of clause (b) of Theorem 3.1 such that $D^*$ is not computable.*

The present section is entirely devoted to a proof of this lemma. But before we start our long journey through that proof, let us see that this lemma implies the completeness of $\mathbf{Int}^{\circ\!-}$, in the strong form of clause (b) of Theorem 3.1.

Indeed, suppose $K$ is an $\mathbf{Int}^{\circ\!-}$-formula not provable in $\mathbf{Int}^{\circ\!-}$. Let $\mathcal{S}$ be the standardization of $K$. By Lemma 5.2, $\mathbf{Int}^{\circ\!-}$ does not prove $\mathcal{S}$. Then, by our main lemma 7.1, $D^*$ is not computable, where $D$ is the desequentization of $\mathcal{S}$ and $*$ is an interpretation satisfying the complexity condition of clause (b) of Theorem 3.1. But then, by Lemma 6.1, for the same interpretation $*$, $K^*$ is not computable. Done!

Thus, as far as the $\mathbf{Int}^{\circ\!-}$ part of our main Theorem 3.1 is concerned, our only remaining duty is to prove Lemma 7.1. The $\mathbf{Int}^{\succ}$ part of Theorem 3.1 will be taken care of in Section 8.

## 7.1  Main claim

Let us get started with our proof of Lemma 7.1. We pick and fix an arbitrary standard $\mathbf{Int}^{\circ\!-}$-sequent

$$\left.\begin{array}{l} X_1 \circ\!\!-(Y_1 \circ\!\!- Z_1),\ \cdots,\ X_s \circ\!\!-(Y_s \circ\!\!- Z_s), \\ (P_1 \circ\!\!- Q_1) \circ\!\!- R_1,\ \cdots,\ (P_s \circ\!\!- Q_s) \circ\!\!- R_s \end{array}\right\} \ \Rightarrow\ W, \tag{8}$$

and assume that $\mathbf{Int}^{\circ\!-} \not\vdash (8)$.

Let us agree for the rest of this section that

*$j$ and $i$ exclusively range over $1,\ldots,s$ and $1,\ldots,2s$, respectively.*

The desequentization of (8) is

$$\begin{array}{l} \downarrow\!\circ(X_1 \wedge Y_1 \to Z_1)\ \wedge\ \cdots\ \wedge\ \downarrow\!\circ(X_s \wedge Y_s \to Z_s)\ \wedge \\ \downarrow\!\circ\big((\downarrow\!\circ P_1 \to Q_1) \to R_1\big)\ \wedge\ \cdots\ \wedge\ \downarrow\!\circ\big((\downarrow\!\circ P_s \to Q_s) \to R_s\big) \end{array} \ \to\ W. \tag{9}$$

Thus, our goal is to find a counterinterpretation (satisfying the complexity condition of Theorem 3.1) for the formula (9), with a *counterinterpretation* here meaning an interpretation $*$ such that $\not\models (9)^*$.

Once again remember, from Section 7 of [10], the distinction between *general* and *elementary* letters. Elementary letters are to be interpreted as predicates (elementary games), while general letters can be interpreted as arbitrary static games. As noted before, the letters (atoms) of both affine logic and intuitionistic logic are general rather than elementary. However, when it comes to interpretations, formulas with only elementary letters — called *elementary-base* formulas — are both technically and intuitively easier to deal with than those with general letters. For this reason, we are going to replace (9) with the elementary-base — though no longer propositional — formula (10) of the same form as (9), and then construct a counterinterpretation for (10) rather than (9).

In particular, for each atom $A$ of the language of $\mathbf{Int}^{\circ\!-}$, we fix a unique 1-ary elementary letter $\dot{A}$ — unique in the sense that whenever $A \neq B$, we also have $\dot{A} \neq \dot{B}$. We also fix a variable $x$ and, for each atom $A$ of the language of $\mathbf{Int}^{\circ\!-}$, agree on the abbreviation $\breve{A}$ defined by

$$\breve{A} \;=\; \sqcup x \dot{A}(x).$$

Now, the above-mentioned elementary-base formula (10) is simply obtained from (9) through replacing every atom $A$ by $\breve{A}$:

$$\begin{array}{l} \wedge_\circ(\breve{X}_1 \wedge \breve{Y}_1 \to \breve{Z}_1) \;\wedge\; \cdots \;\wedge\; \wedge_\circ(\breve{X}_s \wedge \breve{Y}_s \to \breve{Z}_s) \;\wedge\; \\ \wedge_\circ\bigl((\wedge_\circ \breve{P}_1 \to \breve{Q}_1) \to \breve{R}_1\bigr) \;\wedge\; \cdots \;\wedge\; \wedge_\circ\bigl((\wedge_\circ \breve{P}_s \to \breve{Q}_s) \to \breve{R}_s\bigr) \end{array} \;\to\; \breve{W}. \qquad (10)$$

We label the following statement (and subsequent similar statements) "claim" rather than "lemma" because it is true only in our particular context, set by the assumption that (8) is not provable in $\mathbf{Int}^{\circ\!-}$.

**Claim 7.2** *There is an interpretation $\star$ with $\not\models (10)^\star$, satisfying the **complexity condition** that $\star$ interprets every elementary atom as a Boolean combination of $\Sigma_1$-predicates.*

Before we attempt to prove Claim 7.2, let us see that it implies the main Lemma 7.1. Remember that our assumption, within the proof of Lemma 7.1, is that $\mathbf{Int}^{\circ\!-} \not\vdash (8)$. According to Claim 7.2 which is based on that assumption, we have $\not\models (10)^\star$, where $\star$ is a certain interpretation sending every elementary atom to a Boolean combination of $\Sigma_1$-predicates. Let now $*$ be the interpretation that sends every atom $A$ of the language of $\mathbf{Int}^{\circ\!-}$ to $\breve{A}^\star$. Obviously then $(9)^* = (10)^\star$, so that $\not\models (9)^*$. And clearly $*$ does satisfy the complexity condition of clause (b) of Theorem 3.1. As (9) is the desequentization of (8), we find Lemma 7.1 proven.

So, the "only" remaining duty within our proof of the main Lemma 7.1 is to prove Claim 7.2. The rest of this section is solely devoted to that task.

## 7.2 Terminology and notation

Note that, since each atom of (10) is to be interpreted as an elementary game, the structure ($\mathbf{Lr}$ component) of the game $(10)^\star$ does not depend on the selection of an interpretation $\star$. This nice property of elementary-base formulas was one of our reasons for choosing to deal with (10) instead of (9). In many contexts, it allows us to terminologically treat (10) as if it was a game, even though, strictly speaking, it is just a formula, and becomes a game only after an interpretation is applied to it. Namely, we can and will unambiguously say "legal run of (10)", meaning "legal run of $(10)^\star$ for some (= every) interpretation $\star$".

We will often need to differentiate between subformulas of (9) or (10) and particular occurrences of such. It should be remembered that the expressions "$P_j$", "$\breve{P}_j$",

"$\breve{X}_j \wedge \breve{Y}_j$", etc. are metaexpressions, denoting subformulas of (9) or (10). As it happens, for different occurrences of subformulas of (9) or (10) we have chosen different metaexpressions, so those occurrences can be safely identified with the corresponding metaexpressions. To avoid possible notational confusions, we will write "$\lfloor P_j \rfloor$", "$\lfloor \breve{P}_j \rfloor$", "$\lfloor \breve{X}_j \wedge \breve{Y}_j \rfloor$", etc. to indicate that we mean **metaexpressions** (= particular occurrences of subformulas) rather than the formulas for which those expressions stand. So, say, when we write $\breve{X}_i = \breve{X}_j$ or $\breve{X}_i = \breve{Q}_j$, we mean that $\breve{X}_i$ and $\breve{X}_j$, or $\breve{X}_i$ and $\breve{Q}_j$, are identical as formulas; on the other hand, by writing $\lfloor \breve{X}_i \rfloor = \lfloor \breve{X}_j \rfloor$ we will mean that the two metaexpressions "$\breve{X}_i$" and "$\breve{X}_j$" are graphically identical, i.e., that $\breve{X}_i$ and $\breve{X}_j$ stand for the same occurrence of the same subformula of (10), which implies that $i = j$. And, as the expressions "$\breve{X}$" and "$\breve{Q}$" are graphically different from each other, we would never have $\lfloor \breve{X}_i \rfloor = \lfloor \breve{Q}_j \rfloor$, no matter what $i$ and $j$ are.

Consider any particular legal position or run $\Gamma$ of (10). Since (10) is the $\rightarrow$-combination of games, every move of $\Gamma$ has the form $1.\alpha$ or $2.\alpha$. Intuitively, $1.\alpha$ means the move $\alpha$ made in the antecedent of (10), and $2.\alpha$ the move $\alpha$ made in the consequent. Correspondingly, we think of $\Gamma$ as consisting of two subruns which, using the notational conventions of Subsection 4.3 of [10], are denoted by $\Gamma^{1.}$ and $\Gamma^{2.}$. We will be referring to $\Gamma^{2.}$ as the $\Gamma$-**residual position of** $\lfloor \breve{W} \rfloor$, because, intuitively, $\Gamma^{2.}$ is what remains of $\Gamma$ after discarding in it everything but the part that constitutes a run in the $\lfloor \breve{W} \rfloor$ component of (10). Note that we wrote $\lfloor \breve{W} \rfloor$ here. Using just $\breve{W}$ instead could have been ambiguous, for $\breve{W}$, as a formula, may (and probably does) have many occurrences in (10), while $\lfloor \breve{W} \rfloor$ refers to the occurrence of that formula in the consequent and only there. Also, we said "position" rather than "run". It is safe to do so because $\Gamma^{2.}$, which has to be a legal run of the game $\breve{W} = \bigsqcup x \dot{W}(x)$ (for otherwise $\Gamma$ would not be a legal run of (10)), contains at most one labmove — namely, it is $\langle \rangle$ or $\langle \top a \rangle$ for some constant $a$.

As for $\Gamma^{1.}$, it is a legal run of the negation of the antecedent of (10) rather than the antecedent itself. That is so because, as we remember, a game $A \rightarrow B$ is defined as $\neg A \vee B$. And this means nothing but that $\neg \Gamma^{1.}$ is a legal run of the antecedent of (10). So, we will be interested in $\neg \Gamma^{1.}$ rather than $\Gamma^{1.}$, because we prefer to see the antecedent of (10) as it is, without a negation. The antecedent of (10), in turn, is a $\wedge$-conjunction, and we think of $\neg \Gamma^{1.}$ as consisting of as many subruns as the number of conjuncts. Namely, each such subrun is $\neg \Gamma^{1.i.}$ for some $i$. If here $i \in \{1, \ldots, s\}$, we call $\neg \Gamma^{1.i.}$ the $\Gamma$-**residual run of** $\lfloor \circ (\breve{X}_i \wedge \breve{Y}_i \rightarrow \breve{Z}_i) \rfloor$, and if $i = s + j$, we call $\neg \Gamma^{1.i.}$ the $\Gamma$-**residual run of** $\lfloor \circ ((\circ \breve{P}_j \rightarrow \breve{Q}_j) \rightarrow \breve{R}_j) \rfloor$. As in the case of $\lfloor \breve{W} \rfloor$, such names correspond to the intuitive meanings of $\neg \Gamma^{1.i.}$. For example, where $1 \leq j \leq s$, $\neg \Gamma^{1.j.}$ can be characterized as the part of $\Gamma$ that constitutes a run in the $\lfloor \circ (\breve{X}_j \wedge \breve{Y}_j \rightarrow \breve{Z}_j) \rfloor$ component. Such a run should be a legal run of the game (represented by) $\circ (\breve{X}_j \wedge \breve{Y}_j \rightarrow \breve{Z}_j)$, for otherwise $\Gamma$ would not be a legal run of (10).

Assume $\Psi$ is the $\Gamma$-residual run of $\lfloor \circ (\breve{X}_j \wedge \breve{Y}_j \rightarrow \breve{Z}_j) \rfloor$. We will be referring to the bitstring tree $Tree^{\circ(\breve{X}_j \wedge \breve{Y}_j \rightarrow \breve{Z}_j)}\langle \Psi \rangle$ (see Subsection 4.6 of [10]) as the $\Gamma$-**residual** $\lfloor \circ (\breve{X}_j \wedge \breve{Y}_j \rightarrow \breve{Z}_j) \rfloor$-**tree**. Intuitively, this is the underlying BT structure of the subrun of $\Gamma$ that is taking place in the $\lfloor \circ (\breve{X}_j \wedge \breve{Y}_j \rightarrow \breve{Z}_j) \rfloor$ component of (10). Then the run $\Psi$ is further thought of as consisting of multiple legal runs of $\breve{X}_j \wedge \breve{Y}_j \rightarrow \breve{Z}_j$, specifically, the run $\Psi^{\preceq w}$ (again, see Subsection 4.6 of [10]) for each complete branch $w$ of the $\Gamma$-residual $\lfloor \raisebox{-0.5ex}{$\circ$}\hspace{-0.5em}\downarrow (\check{X}_j \wedge \check{Y}_j \to \check{Z}_j) \rfloor$-tree. Notice that such a $\Psi^{\preceq w}$, as a legal run of $\check{X}_j \wedge \check{Y}_j \to \check{Z}_j$, would be finite, containing at most 3 labmoves. Hence we can refer to it as "position" rather than "run". We call such a position $\Psi^{\preceq w}$ the **$\Gamma$-residual position of** $\lfloor \check{X}_j \wedge \check{Y}_j \to \check{Z}_j \rfloor^w$. The **$\Gamma$-residual** $\lfloor \raisebox{-0.5ex}{$\circ$}\hspace{-0.5em}\downarrow ((\raisebox{-0.5ex}{$\circ$}\hspace{-0.5em}\downarrow \check{P}_j \to \check{Q}_j) \to \check{R}_j) \rfloor$**-tree** and the **$\Gamma$-residual run of** $\lfloor (\raisebox{-0.5ex}{$\circ$}\hspace{-0.5em}\downarrow \check{P}_j \to \check{Q}_j) \to \check{R}_j \rfloor^w$ (where $w$ is a complete branch of that tree) are defined similarly. In this case, for safety, we say *run* rather than *position*, for the game (represented by) $(\raisebox{-0.5ex}{$\circ$}\hspace{-0.5em}\downarrow \check{P}_j \to \check{Q}_j) \to \check{R}_j$ is not finite-depth because of its $\raisebox{-0.5ex}{$\circ$}\hspace{-0.5em}\downarrow \check{P}_j$-component, and hence the corresponding $\Psi^{\preceq w}$ may be infinite.

Assume $w$ is a complete branch of the $\Gamma$-residual $\lfloor \raisebox{-0.5ex}{$\circ$}\hspace{-0.5em}\downarrow (\check{X}_j \wedge \check{Y}_j \to \check{Z}_j) \rfloor$-tree, and $\Theta$ is the $\Gamma$-residual position of $\lfloor \check{X}_j \wedge \check{Y}_j \to \check{Z}_j \rfloor^w$. Such a $\Theta$ is thought of as consisting of two subpositions: $\neg \Theta^{1.}$ and $\Theta^{2.}$. We respectively refer to these as the **$\Gamma$-residual position of** $\lfloor \check{X}_j \wedge \check{Y}_j \rfloor^w$ and the **$\Gamma$-residual position of** $\lfloor \check{Z}_j \rfloor^w$. In turn, $\neg \Theta^{1.}$ is further seen as consisting of two subpositions $\neg \Theta^{1.1.}$ and $\neg \Theta^{1.2.}$, to which we respectively refer as the **$\Gamma$-residual position of** $\lfloor \check{X}_j \rfloor^w$ and the **$\Gamma$-residual position of** $\lfloor \check{Y}_j \rfloor^w$. Similarly, if $w$ is a complete branch of the $\Gamma$-residual $\lfloor \raisebox{-0.5ex}{$\circ$}\hspace{-0.5em}\downarrow ((\raisebox{-0.5ex}{$\circ$}\hspace{-0.5em}\downarrow \check{P}_j \to \check{Q}_j) \to \check{R}_j) \rfloor$-tree and $\Theta$ is the $\Gamma$-residual run of $\lfloor (\raisebox{-0.5ex}{$\circ$}\hspace{-0.5em}\downarrow \check{P}_j \to \check{Q}_j) \to \check{R}_j \rfloor^w$, we respectively refer to $\neg \Theta^{1.}, \Theta^{2.}, \Theta^{1.1.}$ and $\neg \Theta^{1.2.}$ as the **$\Gamma$-residual runs of** $\lfloor \raisebox{-0.5ex}{$\circ$}\hspace{-0.5em}\downarrow \check{P}_j \to \check{Q}_j \rfloor^w$, $\lfloor \check{R}_j \rfloor^w$, $\lfloor \raisebox{-0.5ex}{$\circ$}\hspace{-0.5em}\downarrow \check{P}_j \rfloor^w$ and $\lfloor \check{Q}_j \rfloor^w$, respectively (in the cases of $\lfloor \check{R}_j \rfloor^w$ and $\lfloor \check{Q}_j \rfloor^w$ we can always say "position" instead of "run", of course).

Assume $w$ is a complete branch of the $\Gamma$-residual $\lfloor \raisebox{-0.5ex}{$\circ$}\hspace{-0.5em}\downarrow ((\raisebox{-0.5ex}{$\circ$}\hspace{-0.5em}\downarrow \check{P}_j \to \check{Q}_j) \to \check{R}_j) \rfloor$-tree, and $\Upsilon$ is the $\Gamma$-residual run of $\lfloor \raisebox{-0.5ex}{$\circ$}\hspace{-0.5em}\downarrow \check{P}_j \rfloor^w$. We call $Tree^{\raisebox{-0.5ex}{$\circ$}\hspace{-0.5em}\downarrow \check{P}_j}\langle \Upsilon \rangle$ the **$\Gamma$-residual** $\lfloor \raisebox{-0.5ex}{$\circ$}\hspace{-0.5em}\downarrow \check{P}_j \rfloor^w$**-tree**. $\Upsilon$ is then further seen as consisting of multiple legal positions of $\check{P}_j$, specifically, the position $\Upsilon^{\preceq u}$ for each complete branch $u$ of the $\Gamma$-residual $\lfloor \raisebox{-0.5ex}{$\circ$}\hspace{-0.5em}\downarrow \check{P}_j \rfloor^w$-tree. We refer to such a position $\Upsilon^{\preceq u}$ as the **$\Gamma$-residual position of** $\lfloor \check{P}_j \rfloor^w_u$.

In the above terminological conventions we have started writing "$\lfloor \check{X}_i \wedge \check{Y}_i \rfloor^w$", "$\lfloor \check{P}_i \rfloor^w_u$", etc. Formally these, just like simply "$\lfloor \check{X}_i \wedge \check{Y}_i \rfloor$" or "$\lfloor \check{P}_i \rfloor$", are *metaexpressions*. If, say, we write $\lfloor \check{X}_i \rfloor^w = \lfloor \check{X}_j \rfloor^u$, we imply that the two components are graphically the same, here meaning that $i = j$ and $w = u$. Note that such metaexpressions would not always be finite. For instance, $\lfloor \check{X}_i \rfloor^w$ would be infinite if the bitstring $w$ is so; this, however, can only be the case when $\Gamma$ is an infinite run.

What we call the **residual molecules of** $\Gamma$, or simply **$\Gamma$-molecules**, are the following metaexpressions:

- $\lfloor \check{W} \rfloor$;

- $\lfloor \check{X}_j \rfloor^w$, $\lfloor \check{Y}_j \rfloor^w$ and $\lfloor \check{Z}_j \rfloor^w$ for each $j$ and each complete branch $w$ of the $\Gamma$-residual $\lfloor \raisebox{-0.5ex}{$\circ$}\hspace{-0.5em}\downarrow (\check{X}_j \wedge \check{Y}_j \to \check{Z}_j) \rfloor$-tree.

- $\lfloor \check{Q}_j \rfloor^w$ and $\lfloor \check{R}_j \rfloor^w$ for each $j$ and each complete branch $w$ of the $\Gamma$-residual $\lfloor \raisebox{-0.5ex}{$\circ$}\hspace{-0.5em}\downarrow ((\raisebox{-0.5ex}{$\circ$}\hspace{-0.5em}\downarrow \check{P}_j \to \check{Q}_j) \to \check{R}_j) \rfloor$-tree.

- $\lfloor \check{P}_j \rfloor^w_u$ for each $j$, each complete branch $w$ of the $\Gamma$-residual $\lfloor \raisebox{-0.5ex}{$\circ$}\hspace{-0.5em}\downarrow ((\raisebox{-0.5ex}{$\circ$}\hspace{-0.5em}\downarrow \check{P}_j \to \check{Q}_j) \to \check{R}_j) \rfloor$-tree and each complete branch $u$ of the $\Gamma$-residual $\lfloor \raisebox{-0.5ex}{$\circ$}\hspace{-0.5em}\downarrow \check{P}_j \rfloor^w$-tree.

We may say just "**molecule**" instead of "Γ-molecule" when Γ is fixed in a given context or is irrelevant.

We say that the **types** of the above 7 sorts of molecules are $W$, $X_j$, $Y_j$, $Z_j$, $Q_j$, $R_j$ and $P_j$, respectively. When irrelevant, the subscript $j$ can be omitted here. We differentiate between types and what we call **metatypes**. The metatype of $\lfloor \breve{W} \rfloor$ is the metaexpression $\lfloor W \rfloor$; the metatype of $\lfloor \breve{P}_j \rfloor_u^w$ is the metaexpression $\lfloor P_j \rfloor$; the metatype of $\lfloor \breve{X}_j \rfloor^w$ is the metaexpression $\lfloor X_j \rfloor$, and similarly for $\lfloor \breve{Y}_j \rfloor^w$, $\lfloor \breve{Z}_j \rfloor^w$, $\lfloor \breve{Q}_j \rfloor^w$, $\lfloor \breve{R}_j \rfloor^w$. As in the case of types, the subscripts $j$ can be omitted here when irrelevant. Notice that if two molecules have different types, then they also have different metatypes, but not vice versa: for instance, $\lfloor \breve{Q}_j \rfloor^w$ and $\lfloor \breve{R}_j \rfloor^v$ would always have different metatypes (because "$Q$"≠"$R$"), but their types may be identical, meaning that so are the atoms of (9) for the occurrences of which $Q_j$ and $R_j$ stand.

$\lfloor W \rfloor$-, $\lfloor X \rfloor$-, $\lfloor Y \rfloor$- and $\lfloor Q \rfloor$-metatype molecules are said to be **positive**; and $\lfloor Z \rfloor$-, $\lfloor R \rfloor$- and $\lfloor P \rfloor$-metatype molecules are said to be **negative**. When one of two molecules $M_1, M_2$ is positive and the other is negative, we say that $M_1$ and $M_2$ have **opposite genders**. Mark the fact that, for a Γ-residual molecule $M$, the Γ-residual position of $M$ is always $\langle \rangle$ or $\langle \top a \rangle$ for some constant $a$. Note that here, if $M$ is negative, the move $a$ is made by player $\bot$ even though it is $\top$-labeled in $\langle \top a \rangle$.

Let $M$ be a Γ-molecule of type $A \in \{W, X_j, Y_j, Z_j, Q_j, R_j, P_j \mid 1 \leq j \leq s\}$. We say that $M$, *as a Γ-molecule* (or *in Γ*), is:

- **virgin** iff the Γ-residual position of $M$ is empty; in this case we define the **content** of $M$ to be the formula $\breve{A}$, i.e. $\bigsqcup x \dot{A}(x)$;

- **devirginized** iff the Γ-residual position of $M$ is $\langle \top a \rangle$; in this case we define the **content** of $M$ to be the formula $\dot{A}(a)$.

Note that two molecules may have identical contents even if their metatypes (but not types!) are different.

Intuitively, the content of a given Γ-molecule $M$ is the game to which the corresponding subgame of (10) has evolved as a result of the moves of Γ made within the $M$ component. Here *the corresponding (sub)game* is what we see in the expression $M$ between "$\lfloor$" and "$\rfloor$". For instance, the game corresponding to $\lfloor \breve{W} \rfloor$ is $\breve{W}$, the game corresponding to $\lfloor \breve{X}_j \rfloor^w$ is $\breve{X}_j$, and the game corresponding to $\lfloor \breve{P}_j \rfloor_u^w$ is $\breve{P}_j$. For $\lfloor \breve{W} \rfloor$ (and similarly for other molecules), virginity thus means that the moves of Γ have not affected this component (that is, none of those moves were made within $\lfloor \breve{W} \rfloor$), so that, as a (sub)game, it remains $\bigsqcup x \dot{W}(x)$; and being devirginized means that the moves of Γ have brought the game $\bigsqcup x \dot{W}(x)$ down to $\dot{W}(a)$ for some constant $a$.

We say that a Γ-molecule $M_1$ is **matchingly devirginized** (in Γ) iff $M_1$ is devirginized and there is another devirginized Γ-molecule $M_2$ such that $M_1$ and $M_2$ have opposite genders but identical contents. If $M_1$ is devirginized but not matchingly so, then we say that it is **non-matchingly devirginized**.

Note that, for a Γ-residual molecule $M$, the content of $M$, as well as whether $M$ virgin, devirginized or matchingly devirginized, depends on Γ. That is why, unless Γ is fixed or clear from the context, for safety we should say "the content of $M$ in Γ" instead of just "the content of $M$", say "$M$ is devirginized in $\Gamma$" or "$M$, as a $\Gamma$-molecule, is devirginized" instead of just "$M$ is devirginized", etc. The point is that the same metaexpression $M$ can be a residual molecule of two different runs. Relevant to our interests are only the cases when one run, say $\Gamma_1$, is an initial segment of the other run, say $\Gamma_2$. Assume this is so for the rest of the present paragraph, and assume $M$ is a $\Gamma_1$-molecule. If $M = \lfloor \check{W} \rfloor$, $M$ will also be a $\Gamma_2$-molecule. It is possible, however, that $M$ is virgin in $\Gamma_1$ while devirginized in $\Gamma_2$; and if so, the content of $M$ in $\Gamma_1$ will be different from that in $\Gamma_2$. And it is also generally possible that $M$ is non-matchingly devirginized in $\Gamma_1$ while matchingly devirginized in $\Gamma_2$. Similarly when the metatype of $M$ is anything other than $\lfloor W \rfloor$. However, when $M$ is a non-$\lfloor W \rfloor$-metatype $\Gamma_1$-molecule such as, say, $\lfloor \check{X}_j \rfloor^w$, then there is no guarantee that $M$ is also a $\Gamma_2$-molecule. For instance, if ($\Gamma_2 \neq \Gamma_1$ and) in position $\Gamma_1$ the player $\top$ made a replicative move in the $\lfloor \circ^{\downarrow}(\check{X}_j \wedge \check{Y}_j \to \check{Z}_j) \rfloor$ component of the game which split (i.e. extended to $w0$ and $w1$) the leaf $w$ of the $\Gamma_1$-residual $\lfloor \circ^{\downarrow}(\check{X}_j \wedge \check{Y}_j \to \check{Z}_j) \rfloor$-tree, then $w$ would be just an internal node rather than a complete branch of the $\Gamma_2$-residual $\lfloor \circ^{\downarrow}(\check{X}_j \wedge \check{Y}_j \to \check{Z}_j) \rfloor$-tree, meaning that $M = \lfloor \check{X}_j \rfloor^w$ is not a residual molecule of $\Gamma_2$. Instead of $\lfloor \check{X}_j \rfloor^w$, in the general case, $\Gamma_2$ could have many residual molecules of the form $\lfloor \check{X}_j \rfloor^{w'}$, where $w \preceq w'$. Let us agree to say about each such $\Gamma_2$-molecule $\lfloor \check{X}_j \rfloor^{w'}$ that it **descends** from $\lfloor \check{X}_j \rfloor^w$, or that $\lfloor \check{X}_j \rfloor^w$ is **the $\Gamma_1$-predecessor** — or simply **a predecessor** if we do not care about details — of $\lfloor \check{X}_j \rfloor^{w'}$. Similarly for the cases when $M$ is $\lfloor \check{Y}_j \rfloor^w$, $\lfloor \check{Z}_j \rfloor^w$, $\lfloor \check{Q}_j \rfloor^w$ or $\lfloor \check{R}_j \rfloor^w$. And rather similarly for the case $M = \lfloor \check{P}_j \rfloor^w_u$: in this case $M$ will be said to be the $\Gamma_1$-**predecessor** (or just **a predecessor**) of every $\Gamma_2$-molecule $\lfloor \check{P}_j \rfloor^{w'}_{u'}$ such that $w \preceq w'$ and $u \preceq u'$; and, correspondingly, every such $\lfloor \check{P}_j \rfloor^{w'}_{u'}$ will be said to **descend** from $\lfloor \check{P}_j \rfloor^w_u$. Extending this terminology to the remaining case of $M = \lfloor \check{W} \rfloor$, the latter is always its own (single) **descendant** and **predecessor**.

Here comes some more terminology. In the context of a given legal position $\Gamma$ of (10), where $M$ is a virgin $\Gamma$-molecule, to **devirginize** $M$ intuitively means to make a (legal) move that makes $M$ devirginized; we say that such a move is **patient** if $M$ is the only molecule whose content it modifies. Informally speaking, a patient move for a non-$\lfloor W \rfloor$-metatype molecule $M$ means that the move is made in the corresponding leaf (or two nested leaves if $M$ is $\lfloor P \rfloor$-metatype) rather than internal node(s) of the corresponding underlying BT(s), for a move in an internal node $v$ would simultaneously affect several molecules — all those that are associated with leaves $r$ such that $v \preceq r$. In precise terms, we have:

- To devirginize $\lfloor \check{W} \rfloor$ means to make the move $2.a$ for some constant $a$. This sort of a move is automatically patient.

- To devirginize $\lfloor \check{X}_j \rfloor^w$ (resp. $\lfloor \check{Y}_j \rfloor^w$, resp. $\lfloor \check{Z}_j \rfloor^w$) means to make the move $1.j.w'.1.1.a$ (resp. $1.j.w'.1.2.a$, resp. $1.j.w'.2.a$) for some bitstring $w' \preceq w$ and some constant $a$. Such a move is patient iff $w' = w$.

- To devirginize $\lfloor \check{Q}_j \rfloor^w$ (resp. $\lfloor \check{R}_j \rfloor^w$) means to make the move $1.i.w'.1.2.a$ (resp. $1.i.w'.2.a$) for $i = s+j$, some bitstring $w' \preceq w$ and some constant $a$. Again, such a move is patient iff $w' = w$.

- To devirginize $\lfloor \breve{P}_j \rfloor_u^w$ means to make the move $1.i.w'.1.1.u'.a$ for $i = s + j$, some bitstrings $w' \preceq w$ and $u' \preceq u$, and some constant $a$. Such a move patient iff both $w' = w$ and $u' = u$.

Notice that player $\top$ can only devirginize positive molecules, while player $\bot$ can only devirginize negative molecules. Every devirginizing move thus has the form $\alpha.a$ for some constant $a$. Let us call such a constant $a$ *the choice constant* of the devirginizing move. Then we say that a given act (move) of devirginization is done **diversifyingly** iff the move is patient, and its choice constant is the smallest constant that has never been used before in the play (run) as the choice constant of some devirginizing move.

## 7.3 The counterstrategy $\mathcal{E}$

In this subsection we set up a counterstrategy for (10) in the form of an EPM $\mathcal{E}$, which will act in the role of $\bot$ in a play over (10). $\mathcal{E}$ is a *universal counterstrategy* for (10), in the sense that, as will be shown later, no HPM wins $(10)^\star$ against this particular, one-for-all EPM for the yet-to-be-constructed interpretation $\star$. Since $\mathcal{E}$ plays as $\bot$ rather than $\top$, we will be interested in the run *cospelled* rather than spelled by any given computation branch of $\mathcal{E}$. That is, in a play, the moves made by $\mathcal{E}$ get the label $\bot$, and the moves made by its adversary get the label $\top$. In our description of $\mathcal{E}$ and the further analysis of its behavior, we will be relying on the **clean environment assumption**. According to it, the adversary of a given agent (the adversary of $\mathcal{E}$ specifically) never makes illegal moves. As pointed out in [10], such an assumption is perfectly safe and legitimate for, if the adversary makes an illegal move, then $\mathcal{E}$ will be the winner no matter what happens afterwards, and making $\mathcal{E}$ the winner is the very purpose of our construction. From the definition of $\mathcal{E}$ it will be also immediately clear that $\mathcal{E}$ itself does not make any illegal moves either. Since all runs that $\mathcal{E}$ generates are thus legal, we usually omit the word "legal", and by a run or position we will always mean a legal run or position of (10).

The work of $\mathcal{E}$ consists in sequentially performing the first two or all three (depending on how things evolve) of the following procedures FIRST, SECOND and THIRD. In the descriptions of these procedures, "current" should be understood as $\Phi$-residual, where $\Phi$ is the position of the play at the time when a given step is performed. This word may be omitted, and by just saying "molecule" we mean current molecule. Similarly, "virgin", "(matchingly) devirginized", "devirginize", etc. should be understood in the context of the then-current position. Also, since the current position is always a position (finite run), it is safe to say "leaf" instead of "complete branch" when talking about bitstring trees in our description of the work of $\mathcal{E}$.

**PROCEDURE** FIRST: Diversifyingly devirginize all $\lfloor P \rfloor$-metatype molecules, and go to SECOND.

**PROCEDURE** SECOND: If $\lfloor \breve{W} \rfloor$ is matchingly devirginized, go to THIRD. Else perform each of the following routines:
**Routine 1.** For each $j$ and each leaf $w$ of the current $\lfloor \dot{\circ}(\breve{X}_j \wedge \breve{Y}_j \to \breve{Z}_j) \rfloor$-tree, whenever both $\lfloor \breve{X}_j \rfloor^w$ and $\lfloor \breve{Y}_j \rfloor^w$ are matchingly devirginized and $\lfloor \breve{Z}_j \rfloor^w$ is virgin, diversifyingly devirginize $\lfloor \breve{Z}_j \rfloor^w$.

**Routine 2.** For each $j$ and each leaf $w$ of the current $\lfloor \diamondsuit((\diamondsuit \breve{P}_j \to \breve{Q}_j) \to \breve{R}_j) \rfloor$-tree, whenever $\lfloor \breve{Q}_j \rfloor^w$ is matchingly devirginized and $\lfloor \breve{R}_j \rfloor^w$ is virgin, diversifyingly devirginize $\lfloor \breve{R}_j \rfloor^w$.

**Routine 3.** Grant permission, and repeat SECOND.

**PROCEDURE** THIRD: Diversifyingly devirginize all $\lfloor Z \rfloor$- and $\lfloor R \rfloor$-metatype virgin molecules, and go into an infinite loop within a permission state.

Remember that a fair EPM is one whose every $e$-computation branch (any valuation $e$) is fair, i.e. permission is granted infinitely many times in each branch. Before we go any further, let us make the straightforward observation that

$$\mathcal{E} \text{ is a fair EPM.} \tag{11}$$

This is so because THIRD grants permission infinitely many times, and if THIRD is never reached by $\mathcal{E}$, then SECOND will be iterated infinitely many times, with each iteration granting permission in Routine 3.

Our ultimate goal is to show that (10) is *not valid*, which, as mentioned, will be achieved by finding an interpretation $\star$ such that no HPM wins $(10)^\star$ against $\mathcal{E}$. We approach this goal by first proving the weaker fact that (10) is *not uniformly valid*. In particular, below we are going to show that, for any valuation $e$ and any $e$-computation branch $B$ of $\mathcal{E}$, there is an interpretation $\dagger$ such that the run cospelled by $B$ is a $\perp$-won run of $(10)^\dagger$. As will be observed at the beginning of Subsection 7.6, this fact immediately implies the non-uniform-validity of (10). Such a $B$-depending counterinterpretation is going to be what in [8] was called **perfect**, in our particular case meaning that for any predicate letter $\dot{A}$ of (10), $\dot{A}^\dagger(x)$ is a finitary predicate that does not depend on any variables except $x$. This can be easily seen to make $\breve{A}^\dagger$ and hence $(10)^\dagger$ a constant game, allowing us to safely ignore the valuation parameter $e$ in most contexts, which is irrelevant because neither the game $e[\breve{A}^\dagger]$ nor (notice) the work of $\mathcal{E}$ depends on $e$. To define such an interpretation, it is sufficient to indicate what constant atomic formulas are made by it true and what constant atomic formulas are made false. Here and later, for simplicity, by "constant atomic formulas" we mean formulas of the form $\dot{A}(a)$, where $\dot{A}$ is a predicate letter occurring in (10) and $a$ is a constant. For obvious reasons, how $\dagger$ interprets any other atoms is irrelevant, and we may safely pretend that such atoms simply do not exist in the language.

So, fix any valuation $e$ spelled on the valuation tape of $\mathcal{E}$, and any $e$-computation branch $B$ of $\mathcal{E}$. Let $\Gamma$ be the run cospelled by $B$. Let us agree to say "**ultimate**" (run, position, tree) for "$\Gamma$-residual" (run, position, tree). To $\Gamma$ itself we refer as the **ultimate run of** $\lfloor (10) \rfloor$, or simply the **ultimate run**. By just saying "**molecule**" we mean a residual molecule of $\Gamma$ or of any initial segment of it. And $\Gamma$-molecules we call **ultimate molecules**. Any molecule would thus be an ultimate molecule or a predecessor of such. When $M$ is an ultimate molecule, by just saying that $M$ is virgin, devirginized or matchingly devirginized we mean that $M$ is so in $\Gamma$.

We will say that the branch $B$ is **short** iff, in the process of playing it up, $\mathcal{E}$ never entered the THIRD stage, thus forever remaining in SECOND. Otherwise $B$ is **long**. The scope of all this $B$- and $\Gamma$-dependent terminology extends to the following two subsections, throughout which $B$ and $\Gamma$ are fixed.

Our construction of a counterinterpretation for (10) depends on whether $B$ is short or long. We consider these two cases separately.

## 7.4 Constructing a counterinterpretation when $B$ is short

Assume $B$ is short. This, looking at the first line of the description of SECOND, means that, in the ultimate run of $\lfloor(10)\rfloor$ (in $\Gamma$, that is), $\breve{W}$ is not matchingly devirginized. We select our

$$counterinterpretation \quad ^\dagger$$

to be the perfect interpretation that makes the contents of all positive non-matchingly devirginized ultimate molecules false, and all other constant atomic formulas true.

**Convention 7.3**

- Where $\dot{A}$ is a predicate letter of (10) and $a$ is any constant, we say that $\dot{A}(a)$ is **true** iff $\dot{A}^\dagger(a)$ is true.

- We say that $\lfloor(10)\rfloor$ is **true** iff the ultimate run of $\lfloor(10)\rfloor$ is a $\top$-won run of $(10)^\dagger$.

- We say that $\lfloor\breve{W}\rfloor$ is **true** iff the ultimate position of $\lfloor\breve{W}\rfloor$ is a $\top$-won position of $\breve{W}^\dagger$.

- We say that $\lfloor\dot{\circ}(\breve{X}_j \wedge \breve{Y}_j \to \breve{Z}_j)\rfloor$ is **true** iff the ultimate run of $\lfloor\dot{\circ}(\breve{X}_j \wedge \breve{Y}_j \to \breve{Z}_j)\rfloor$ is a $\top$-won run of $\bigl(\dot{\circ}(\breve{X}_j \wedge \breve{Y}_j \to \breve{Z}_j)\bigr)^\dagger$. Similarly for $\lfloor\dot{\circ}\bigl((\dot{\circ}\breve{P}_j \to \breve{Q}_j) \to \breve{R}_j\bigr)\rfloor$.

- Where $w$ is a complete branch of the ultimate $\lfloor\dot{\circ}(\breve{X}_j \wedge \breve{Y}_j \to \breve{Z}_j)\rfloor$-tree, we say that $\lfloor\breve{Z}_j\rfloor^w$ is **true** iff the ultimate run (position) of $\lfloor\breve{Z}_j\rfloor^w$ is a $\top$-won run of $\breve{Z}_j^\dagger$. Similarly for $\lfloor\breve{X}_j\rfloor^w$, $\lfloor\breve{Y}_j\rfloor^w$, $\lfloor\breve{X}_j \wedge \breve{Y}_j\rfloor^w$, $\lfloor\breve{Q}_j\rfloor^w$, $\lfloor\breve{R}_j\rfloor^w$, $\lfloor\dot{\circ}\breve{P}_j \to \breve{Q}_j\rfloor^w$.

- Where $w$ is a complete branch of the ultimate $\lfloor\dot{\circ}\bigl((\dot{\circ}\breve{P}_j \to \breve{Q}_j) \to \breve{Z}_j\bigr)\rfloor$-tree and $u$ is a complete branch of the ultimate $\lfloor\dot{\circ}\breve{P}_j\rfloor^w$-tree, we say that $\lfloor\breve{P}_j\rfloor^w_u$ is **true** iff the ultimate position of $\lfloor\breve{P}_j\rfloor^w_u$ is a $\top$-won position of $\breve{P}_j^\dagger$.

"**False**" will mean "not true".

Based on the definition of $\sqcup$, with a moment's thought we can see that the following claim is valid:

**Claim 7.4** *Let $M$ be an arbitrary ultimate molecule.*

**(i)** *If $M$ is virgin, then $M$ is false.*

**(ii)** *If $M$ is devirginized, then $M$ is true iff its content is true.*

Our goal is to show that $\lfloor(10)\rfloor$ is false.

Note that $\lfloor \check{W} \rfloor$ is guaranteed to be false. Indeed, if $\lfloor \check{W} \rfloor$ is virgin, it is false by Claim 7.4(i). And, if $\lfloor \check{W} \rfloor$ is devirginized, then, as no switch to THIRD has occured, $\lfloor \check{W} \rfloor$ must be non-matchingly devirginized. Then, with Claim 7.4(ii) in mind, $\lfloor \check{W} \rfloor$ can be seen to be false by our choice of $^\dagger$.

As $\lfloor \check{W} \rfloor$ is false, in order to show that $\lfloor(10)\rfloor$ is false, it would suffice to verify that, for each $j$, both $\lfloor \natural_\circ (\check{X}_j \wedge \check{Y}_j \to \check{Z}_j) \rfloor$ and $\lfloor \natural_\circ ((\natural_\circ \check{P}_j \to \check{Q}_j) \to \check{R}_j) \rfloor$ are true. Why this would suffice can be seen directly from the definitions of $\wedge$ and $\to$. In turn, based on the definition of $\natural_\circ$, the truth of $\lfloor \natural_\circ (\check{X}_j \wedge \check{Y}_j \to \check{Z}_j) \rfloor$ and $\lfloor \natural_\circ ((\natural_\circ \check{P}_j \to \check{Q}_j) \to \check{R}_j) \rfloor$ means nothing but that:

(a) for every complete branch $w$ of the ultimate $\lfloor \natural_\circ (\check{X}_j \wedge \check{Y}_j \to \check{Z}_j) \rfloor$-tree, $\lfloor \check{X}_j \wedge \check{Y}_j \to \check{Z}_j \rfloor^w$ is true, and

(b) for every complete branch $w$ of the ultimate $\lfloor \natural_\circ ((\natural_\circ \check{P}_j \to \check{Q}_j) \to \check{R}_j) \rfloor$-tree, $\lfloor (\natural_\circ \check{P}_j \to \check{Q}_j) \to \check{R}_j \rfloor^w$ is true.

Pick any $j$, and assume $w$ is as in (a). If $\lfloor \check{Z}_j \rfloor^w$ is devirginized, then, with Claim 7.4(ii) in mind, it is true. This is so because, by our choice of $^\dagger$, only the contents of positive non-matchingly devirginized (ultimate) molecules are made false by this interpretation; $\lfloor \check{Z}_j \rfloor^w$ is not positive, nor is its content the same as that of some positive non-matchingly devirginized ultimate molecule, for then that molecule would not be non-matchingly devirginized. The truth of $\lfloor \check{Z}_j \rfloor^w$, in turn, by the definition of $\to$, can be seen to imply the truth of $\lfloor \check{X}_j \wedge \check{Y}_j \to \check{Z}_j \rfloor^w$. Suppose now $\lfloor \check{Z}_j \rfloor^w$ is virgin. Analyzing how $\mathcal{E}$ acts in Routine 1 of SECOND, the fact that $\lfloor \check{Z}_j \rfloor^w$ is not devirginized can be seen to indicate that at least one of the two molecules $\lfloor \check{X}_j \rfloor^w$ and $\lfloor \check{Y}_j \rfloor^w$ — let us assume it is $\lfloor \check{X}_j \rfloor^w$ — is not matchingly devirginized. If $\lfloor \check{X}_j \rfloor^w$ is virgin, it is false as are all virgin molecules. And if $\lfloor \check{X}_j \rfloor^w$ is non-matchingly devirginized, it is again false by our choice of $^\dagger$. In either case $\lfloor \check{X}_j \rfloor^w$ is thus false and hence, as can be seen from the definition of $\wedge$, so is $\lfloor \check{X}_j \wedge \check{Y}_j \rfloor^w$. This, in turn, by the definition of $\to$, makes $\lfloor \check{X}_j \wedge \check{Y}_j \to \check{Z}_j \rfloor^w$ true. Statement (a) is taken care of.

Assume now $w$ is as in (b). If $\lfloor \check{R}_j \rfloor^w$ is devirginized, then we are done for the same reasons as in the previous paragraph when discussing the similar case for $\lfloor \check{Z}_j \rfloor^w$. Suppose now $\lfloor \check{R}_j \rfloor^w$ is virgin. Reasoning as we did in the corresponding case of the previous paragraph for $\lfloor \check{X}_j \rfloor^w$ (only appealing to Routine 2 of SECOND instead of Routine 1), we find that $\lfloor \check{Q}_j \rfloor^w$ is false. Therefore, in order to conclude that $\lfloor (\natural_\circ \check{P}_j \to \check{Q}_j) \to \check{R}_j \rfloor^w$ is true, it remains to show that $\lfloor \natural_\circ \check{P}_j \rfloor^w$ is true. The truth of $\lfloor \natural_\circ \check{P}_j \rfloor^w$ means nothing but that, for every complete branch $u$ of the ultimate $\lfloor \natural_\circ \check{P}_j \rfloor^w$-tree, $\lfloor \check{P}_j \rfloor^w_u$ is true. But this is indeed so. Consider any complete branch $u$ of the ultimate $\lfloor \natural_\circ \check{P}_j \rfloor^w$-tree. $\lfloor \check{P}_j \rfloor^w_u$ is devirginized due to the actions of $\mathcal{E}$ during FIRST. And, as this molecule is negative, it can be seen to be true by our choice of $^\dagger$. Statement (b) is thus also taken care of.

So, $\lfloor(10)\rfloor$ is false, meaning that $\Gamma$ is a $\bot$-won run of $(10)^\dagger$, i.e. $\mathbf{Wn}^{(10)^\dagger}\langle\Gamma\rangle = \bot$.

## 7.5 Constructing a counterinterpretation when $B$ is long

Throughout this subsection, we assume $B$ is long, meaning that $\lfloor \breve{W} \rfloor$ is matchingly devirginized and thus, at some point of playing $B$ up, $\mathcal{E}$ switched from SECOND to THIRD. We let $\Delta$ denote the (finite) initial segment of our ultimate run $\Gamma$ consisting of all of the (lab)moves made before this switch to THIRD occurred.

By a **supermolecule** we mean a devirginized residual molecule $M$ of some subposition (i.e. a finite initial segment) $\Phi$ of $\Gamma$ such that for no proper initial segment $\Psi$ of $\Phi$ is the $\Psi$-predecessor of $M$ devirginized. We refer to such a $\Phi$ as the **position of devirginization** of $M$, and refer to the length of $\Phi$ as the **time of devirginization** of $M$. Next, if here $\Phi$ is not longer than $\Delta$, we say that $M$ is an **OGSM** (*Old-Generation SuperMolecule*). Intuitively, OGSMs are molecules that lost their innocence while $\mathcal{E}$ was still acting within FIRST or SECOND. Note that $\lfloor \breve{W} \rfloor$ is an OGSM.

Obviously any devirginized residual molecule $M$ (of $\Gamma$ or any of its initial segments) descends from some unique supermolecule. We call the supermolecule from which $M$ descends the **essence** of $M$. Every supermolecule is thus its own essence. Only devirginized molecules have essences. Therefore, if we say "the essence of $M$", the claim that $M$ is devirginized is automatically implied. Whether a devirginized molecule $M$ is a supermolecule or not, by the **position of devirginization** and **time of devirginization** of $M$ we mean those of the essence of $M$. Intuitively, the position of devirginization of such an $M$ is the position in which $M$ — more precisely, a predecessor of $M$ — first became devirginized, and the time of devirginization tells us how soon after the start of the play this happened. Mark the obvious fact that the content of any devirginized molecule is the same as that of the essence of that molecule.

**Claim 7.5** *No two different negative supermolecules have identical contents.*

**Proof.** This is so because negative molecules are devirginized by $\mathcal{E}$, which always does devirginization in a diversifying way — in a way that ensures that the content of the resulting molecule is different from that of any other devirginized molecule. $\square$

We define a **chain** as any nonempty finite sequence $\langle M_1, \ldots, M_m \rangle$ of OGSMs such that:

1. $M_1$ and only $M_1$ is $\lfloor P \rfloor$-metatype.

2. For each odd $k$ with $1 \leq k < m$, $M_k$ is negative, $M_{k+1}$ is positive, and the two OGSMs have identical contents.

3. For each odd $k$ with $3 \leq k \leq m$, we have:

    - if $M_k = \lfloor \breve{Z}_j \rfloor^w$ (some $j, w$), then $M_{k-1}$ is the essence of $\lfloor \breve{X}_j \rfloor^w$ or the essence of $\lfloor \breve{Y}_j \rfloor^w$;
    - if $M_k = \lfloor \breve{R}_j \rfloor^w$ (some $j, w$), then $M_{k-1}$ is the essence of $\lfloor \breve{Q}_j \rfloor^w$.

As an aside, observe that, according to the above definition, if $\lfloor \breve{W} \rfloor$ is in a chain, then it can only be the last element of the chain, and such a chain is of an even length. All internal (neither the first nor the last) odd-numbered elements of a chain are $\lfloor Z \rfloor$- or $\lfloor R \rfloor$-metatype, and all internal even-numbered elements are $\lfloor X \rfloor$-, $\lfloor Y \rfloor$- or $\lfloor Q \rfloor$-metatype.

We say that a chain $\langle M_1, \ldots, M_m \rangle$ is **open** iff, where $\lfloor P_j \rfloor$ is the metatype of $M_1$, no element of the chain has the metatype $\lfloor Q_j \rfloor$ (for the same $j$).

For an OGSM $M$, by $\mathbf{Base}(M)$ we denote the set of all atoms $P_j$ of (9) such that there is an open chain $\langle M_1, \ldots, M_m \rangle$ with $M_1$ being $P_j$-type and $M_m = M$.

Now remember our assumption that $\mathbf{Int}^{\circ\!\!-}$ does not prove (8). By the completeness with respect to Kripke semantics, this means that there is a tree-like Kripke model $(\mathcal{W}, \mathcal{R}, \models)$ such that (8) is false in the root world *root* of it. Obviously here we may assume that every formula of the antecedent of (8) is true in *root*, while $W$ is false. We consider this model and its root world *root* fixed throughout this section.

**Claim 7.6** *Suppose $M$ is a negative $A$-type ($A \in \{P_j, R_j, Z_j \mid 1 \leq j \leq s\}$) OGSM, and $p$ is a world with $p \models \mathbf{Base}(M)$. Then $p \models A$.*

**Proof.** Let $M$, $A$, $p$ be as above, and let $\Phi$ be the position of devirginization of $M$. We proceed by induction on the time of devirginization of $M$, i.e. the length of $\Phi$. There are three cases to consider, depending on the metatype of $M$.

*Case 1:* $M$ is $\lfloor P \rfloor$-metatype. Then $\mathbf{Base}(M) = \{A\}$, and hence this case is trivial.

*Case 2:* $M$ is $\lfloor Z \rfloor$-metatype, specifically, $M = \lfloor \check{Z}_j \rfloor^w$. As every formula of the antecedent of (8) is true in *root*, those formulas remain true in all worlds. In particular, $p \models X_j \circ\!\!- (Y_j \circ\!\!- Z_j)$. Hence, in order to show that $p \models Z_j$, it would be sufficient to verify that $p \models X_j$ and $p \models Y_j$. From an analysis of Routine 1 of SECOND it can be seen that the $\Phi$-molecule $\lfloor \check{X}_j \rfloor^w$ is matchingly devirginized, for otherwise $\mathcal{E}$ would not have devirginized $\lfloor \check{Z}_j \rfloor^w$. So, there is a negative $\Phi$-molecule $N$ with the same content as $\lfloor \check{X}_j \rfloor^w$. The time of devirginization of $N$ is, of course, smaller than that of $M$, because $(\lfloor \check{X}_j \rfloor^w$ and hence) $N$ would have to be already devirginized by the time when $M$ lost innocence. Let $K$ be the essence of $N$. As ($N$ and hence) $K$ has the same content as $\lfloor \check{X}_j \rfloor^w$, the type of $K$ should be $X_j$. Now, notice that $\mathbf{Base}(K) \subseteq \mathbf{Base}(M)$. This is so because every open chain $\langle M_1, \ldots, M_m \rangle$ with $M_m = K$ remains an open chain after adding to it the essence of $\lfloor \check{X}_j \rfloor^w$ and then $M$. Hence our assumption $p \models \mathbf{Base}(M)$ implies $p \models \mathbf{Base}(K)$. And, since the time of devirginization of ($N$ and hence of) $K$ is smaller that of $M$, the induction hypothesis yields $p \models X_j$. A similar argument shows that we also have $p \models Y_j$.

*Case 3:* $M$ is $\lfloor R \rfloor$-metatype, specifically, $M = \lfloor \check{R}_j \rfloor^w$. We want to show that $p \models P_j \circ\!\!- Q_j$, from which $p \models R_j$ follows for reasons similar to those pointed out in the previous case. For a contradiction, assume $p \not\models P_j \circ\!\!- Q_j$. Then there is a world $q$ accessible from $p$ such that $q \models P_j$ and $q \not\models Q_j$. From Routine 2 of SECOND we see that the $\Phi$-molecule $\lfloor \check{Q}_j \rfloor^w$ should be matchingly devirginized (otherwise $\mathcal{E}$ would not have devirginized $M$). Arguing as we did in Case 2 for $\lfloor \check{X}_j \rfloor^w$, we find that there is a negative OGSM $K$ whose type is $Q_j$ and whose time of devirginization is smaller than that of $M$. Every element of $\mathbf{Base}(K)$, except perhaps $P_j$, can be seen to be also an element of $\mathbf{Base}(M)$. Therefore, remembering that $q \models P_j$, in conjunction with our assumption that $p \models \mathbf{Base}(M)$ and hence $q \models \mathbf{Base}(M)$, we find $q \models \mathbf{Base}(K)$. This, by the induction hypothesis, implies $q \models Q_j$, which is a contradiction. □

**Claim 7.7** $\lfloor \breve{W} \rfloor$ *is in some open chain.*

**Proof.** Since $B$ is long, $\lfloor \breve{W} \rfloor$ is matchingly devirginized (already) in $\Delta$. Let $N$ be a negative devirginized $\Delta$-molecule whose content is identical with that of $\lfloor \breve{W} \rfloor$, and let $M$ be the essence of $N$. The type of ($N$ and hence) $M$, of course, should be $W$. Remember that $root \not\models W$. Therefore, by Claim 7.6, $root \not\models \mathbf{Base}(M)$. Consequently, $\mathbf{Base}(M) \neq \emptyset$, for otherwise we would vacuously have $root \models \mathbf{Base}(M)$. The fact that $\mathbf{Base}(M)$ is nonempty means that there is an open chain $\langle M_1, \ldots, M_m \rangle$ with $M_m = M$. The result of adding $\lfloor \breve{W} \rfloor$ to that chain obviously remains an open chain. □

Let us select and fix some — say, lexicographically the smallest — open chain $\langle M_1, \ldots, M_m \rangle$ with $M_m = \lfloor \breve{W} \rfloor$, and call it the **master chain**. According to Claim 7.7, such a chain exists. Let us call the OGSMs that are in the master chain **master OGSMs**.

Now we are ready to define the

$$counterinterpretation \ ^\dagger.$$

We define it as the perfect interpretation that makes the content of each master OGSM false, and makes all other constant atomic formulas true.

In what follows, we fully adopt the terminology of Convention 7.3, with the only difference that now the underlying interpretation $^\dagger$ on which that terminology is based is $^\dagger$ as defined in the present subsection rather than as defined in Subsection 7.4. As in Subsection 7.4, our goal is to show that $\lfloor (10) \rfloor$ is false.

That $\lfloor \breve{W} \rfloor$ is false is immediate from our choice of $^\dagger$. Hence, in order to show that $\lfloor (10) \rfloor$ is false, it would suffice to verify that, for each $j$, we have:

(a) for every complete branch $w$ of the ultimate $\lfloor \downarrow_\circ (\breve{X}_j \wedge \breve{Y}_j \rightarrow \breve{Z}_j) \rfloor$-tree, $\lfloor \breve{X}_j \wedge \breve{Y}_j \rightarrow \breve{Z}_j \rfloor^w$ is true, and

(b) for every complete branch $w$ of the ultimate $\lfloor \downarrow_\circ ((\downarrow_\circ \breve{P}_j \rightarrow \breve{Q}_j) \rightarrow \breve{R}_j) \rfloor$-tree, $\lfloor (\downarrow_\circ \breve{P}_j \rightarrow \breve{Q}_j) \rightarrow \breve{R}_j \rfloor^w$ is true.

Pick any $j$, and assume $w$ is as in (a). $\lfloor \breve{Z}_j \rfloor^w$ is devirginized at least due to the actions of $\mathcal{E}$ during THIRD (in case it otherwise managed to stay virgin throughout SECOND). If the essence of $\lfloor \breve{Z}_j \rfloor^w$ is not in the master chain, then its content is true, because the latter, in view of Claim 7.5 and with a little thought, can be seen to be different from the content of any master OGSM, and, by our choice of $^\dagger$, this interpretation only falsifies the contents of master OGSMs. In turn, from the truth of the content of the essence of $\lfloor \breve{Z}_j \rfloor^w$ and hence the truth of $\lfloor \breve{Z}_j \rfloor^w$ we infer that $\lfloor \breve{X}_j \wedge \breve{Y}_j \rightarrow \breve{Z}_j \rfloor^w$ is true. Suppose now the essence of $\lfloor \breve{Z}_j \rfloor^w$ is in the master chain. Then, from the definition of chain and with some thought, we can see that either the essence of $\lfloor \breve{X}_j \rfloor^w$ or the essence of $\lfloor \breve{Y}_j \rfloor^w$ should be there, too. Specifically, such a molecule would be the one immediately preceeding the essence of $\lfloor Z_j \rfloor^w$ in the master chain. Hence, by our choice of $^\dagger$, either $\lfloor \breve{X}_j \rfloor^w$ or $\lfloor \breve{Y}_j \rfloor^w$ is false, which makes $\lfloor \breve{Z}_j \wedge \breve{Y}_j \to \breve{Z}_j \rfloor^w$ true.

Assume now $w$ is as in (b). As was the case with $\lfloor \breve{Z}_j \rfloor^w$ in the previous paragraph, $\lfloor \breve{R}_j \rfloor^w$ is devirginized, and, if the essence of $\lfloor \breve{R}_j \rfloor^w$ is not in the master chain, then its content and hence $\lfloor (\downarrow\!\!\circ \breve{P}_j \to \breve{Q}_j) \to \breve{R}_j \rfloor^w$ is true. Suppose now the essence of $\lfloor \breve{R}_j \rfloor^w$ is in the master chain. Arguing as we did in the previous case for $\lfloor \breve{X}_j \rfloor^w$ or $\lfloor \breve{Y}_j \rfloor^w$, we find that the essence of $\lfloor \breve{Q}_j \rfloor^w$ is a master OGSM and hence $\lfloor \breve{Q}_j \rfloor^w$ is false. Therefore, in order to conclude that $\lfloor (\downarrow\!\!\circ \breve{P}_j \to \breve{Q}_j) \to \breve{R}_j \rfloor^w$ true, it would suffice to show that $\lfloor \downarrow\!\!\circ \breve{P}_j \rfloor^w$ is true. The truth of $\lfloor \downarrow\!\!\circ \breve{P}_j \rfloor^w$ means nothing but that, for every complete branch $u$ of the ultimate $\lfloor \downarrow\!\!\circ \breve{P}_j \rfloor^w$-tree, $\lfloor \breve{P}_j \rfloor^w_u$ is true. But this is indeed so. Consider any complete branch $u$ of the ultimate $\lfloor \downarrow\!\!\circ \breve{P}_j \rfloor^w$-tree. $\lfloor \breve{P}_j \rfloor^w_u$ is devirginized due to the actions during FIRST. Its essence cannot be in the master chain because the essence of $\lfloor \breve{Q}_j \rfloor^w$ is there and the master chain is open. But then, in view of Claim 7.5, the content of $\lfloor \breve{P}_j \rfloor^w_u$ is different from that of any master OGSM, which, by our choice of $^\dagger$, makes $\lfloor \breve{P}_j \rfloor^w_u$ true. Statement (b) is thus also taken care of.

So, $\lfloor (10) \rfloor$ is false, meaning that $\Gamma$ is a $\bot$-won run of $(10)^\dagger$, i.e. $\mathbf{Wn}^{(10)^\dagger}\langle\Gamma\rangle = \bot$.

## 7.6 From non-uniform-validity to non-validity

In the previous two subsections we in fact showed that

$$(10) \text{ is not uniformly valid.} \tag{12}$$

Indeed, suppose, for a contradiction, that (10) is uniformly valid. Let then $\mathcal{H}$ be a uniform solution for (10) — an HPM that wins $(10)^*$ for every interpretation $^*$. Pick an arbitrary valuation $e$ (which is irrelevant here anyway), and let $B$ be the $(\mathcal{E}, e, \mathcal{H})$-branch. As observed in (11), $\mathcal{E}$ is fair, so the conditions of Lemma 4.1 are met and such a branch $B$ is defined. Let then $^\dagger$ be the interpretation constructed from $B$ as in Subsection 7.4 or 7.5, depending on whether $B$ is short or long. Then, as we showed in those two subsections, the run $\Gamma$ cospelled by $B$ is a $\top$-lost run of $(10)^\dagger$ $(= e[(10)^\dagger]$, because the interpretation is perfect). But, by Lemma 4.1, the same $\Gamma$ is a run spelled by some $e$-computation branch of $\mathcal{H}$ — specifically, it is the $\mathcal{H}$ vs. $\mathcal{E}$ run on $e$. This means that $\mathcal{H}$ does not win $(10)^\dagger$, contrary to our assumption that $\mathcal{H}$ is a uniform solution for (10).

In turn, (12) can be eventually rather easily translated into the fact of completeness of $\mathbf{Int}^{\circ\!\!-}$ with respect to uniform validity.

Our goal, however, is to show the completeness of $\mathbf{Int}^{\circ\!\!-}$ with respect to validity rather than just uniform validity — the goal which, as we remember, has been reduced to showing the non-validity of (10). The fact of non-validity of (10), of course, is stronger than the fact of its non-uniform-validity. The point is that the counterinterpretation $^\dagger$ constructed in Subsections 7.4 and 7.5 depends on the computation branch $B$ and hence on the HPM $\mathcal{H}$ that plays against $\mathcal{E}$. That is, different $\mathcal{H}$s could require different $^\dagger$s. In order to show that (10) is not valid, we need to construct a *one*, common-for-all-$\mathcal{H}$ counterinterpretation $^\star$, with the property that every HPM $\mathcal{H}$ loses $(10)^\star$ against $\mathcal{E}$ for that very interpretation $^\star$ on some valuation $e$. Not to worry, we can handle that.

Let us fix the sequence

$$\mathcal{H}_1,\ \mathcal{H}_2,\ \mathcal{H}_3,\ \ldots$$

of all HPMs, enumerated according to the lexicographic order of their standardized descriptions. Next, we select a variable $z$ different from $x$ and, for each constant $c$, define

$$e_c$$

to be the valuation that sends $z$ to $c$, and (arbitrarily) sends all other variables to 1.

For each constant $c$, let

$$B_c$$

be the $(\mathcal{E}, e_c, \mathcal{H}_c)$-branch, and let

$$\Gamma_c$$

be the run cospelled by $B_c$, i.e. (by Lemma 4.1) the $\mathcal{H}_c$ vs. $\mathcal{E}$ run on $e_c$.

Next, for any constant $c$, let

$$\ddagger c$$

be the interpretation constructed from $B_c$ and $\Gamma_c$ in the way we constructed the interpretation $^\dagger$ from $B$ and $\Gamma$ in:

$$\text{Subsection 7.4 if } B_c \text{ is short;}$$
$$\text{Subsection 7.5 if } B_c \text{ is long.}$$

From the results of Subsections 7.4 and 7.5 we thus have:

$$\text{For any constant } c,\ \mathbf{Wn}^{(10)^{\ddagger c}}\langle \Gamma_c\rangle = \bot. \tag{13}$$

Consider any atom $\dot{A}(x)$ of (10). For each particular constant $c$, the predicate $\dot{A}^{\ddagger c}(x)$ is unary, depending only on $x$. But if here $c$ is seen as a variable and, as such, renamed into $z$, then the predicate becomes binary, depending on $x$ and $z$. Let us denote such a predicate by $\ddot{A}(x, z)$. That is, we define $\ddot{A}(x, z)$ as the predicate such that, for any constants $a$ and $c$,

$$\ddot{A}(a, c) = \dot{A}^{\ddagger c}(a).$$

We now define our ultimate

*counterinterpretation* $^\star$

by stipulating that, for any atom $\dot{A}(x)$ of (10),[9] $\dot{A}^\star(x)$ is nothing but the above predicate $\ddot{A}(x, z)$. Note that, unlike $^{\ddagger c}$ (any particular constant $c$), $^\star$ is not a perfect interpretation, for $\dot{A}^\star(x)$ depends on the "hidden" variable $z$.

---

[9]As noted for $^\dagger$ in Subsection 7.3, how $^\star$ interprets the atoms that are not in (10) is irrelevant and not worth bothering at this point.

**Claim 7.8** *For any constant $c$, $(10)^{\ddagger c} = e_c[(10)^\star]$.*

**Proof.** Consider any predicate letter $\dot{A}$ of (10), and any constant $c$. We claim that
$$\left(\bigsqcup x \dot{A}(x)\right)^{\ddagger c} = e_c[\left(\bigsqcup x \dot{A}(x)\right)^\star]. \tag{14}$$

The above two games can be rewritten as $\bigsqcup x \dot{A}^{\ddagger c}(x)$ and $e_c[\bigsqcup x \dot{A}^\star(x)]$, respectively. $e_c[\bigsqcup x \dot{A}^\star(x)]$ can be further rewritten as $e_c[\bigsqcup x \ddot{A}(x,z)]$. Thus, in order to verify (14), we need to show that
$$\bigsqcup x \dot{A}^{\ddagger c}(x) = e_c[\bigsqcup x \ddot{A}(x,z)]. \tag{15}$$

As noted earlier, $\bigsqcup x \dot{A}^{\ddagger c}(x)$ is a constant game due to the fact that the interpretation $\ddagger c$ is perfect. So, it is its own instance, and $\bigsqcup x \dot{A}^{\ddagger c}(x)$ can be safely written instead of $e[\bigsqcup x \dot{A}^{\ddagger c}(x)]$ for whatever (irrelevant) valuation $e$. Of course, the two games of (15) have the same legal runs, with every such run being $\langle\rangle$ or $\langle\top a\rangle$ for some constant $a$. So, we only need to verify that the **Wn** components of those two games are also identical. And, of course, considering only legal runs when comparing the two **Wn** components is sufficient. Furthermore, the empty run is a $\bot$-won run of both games. So, we only need to focus on legal runs of length 1. Consider any such run $\langle\top a\rangle$. By the definition of $\bigsqcup$, $\mathbf{Wn}^{\bigsqcup x \dot{A}^{\ddagger c}(x)}\langle\top a\rangle = \top$ iff $\mathbf{Wn}^{\dot{A}^{\ddagger c}(a)}\langle\rangle = \top$; in turn, $\mathbf{Wn}^{\dot{A}^{\ddagger c}(a)}\langle\rangle = \top$ means nothing but that $\dot{A}^{\ddagger c}(a)$ is true; and, by the definition of $\ddot{A}$, $\dot{A}^{\ddagger c}(a) = \ddot{A}(a,c)$. Thus,
$$\mathbf{Wn}^{\bigsqcup x \dot{A}^{\ddagger c}(x)}\langle\top a\rangle = \top \text{ iff } \ddot{A}(a,c) \text{ is true}. \tag{16}$$

Next, again by the definition of $\bigsqcup$, we have
$$\mathbf{Wn}_{e_c}^{\bigsqcup x \ddot{A}(x,z)}\langle\top a\rangle = \mathbf{Wn}_{e_c}^{\ddot{A}(a,z)}\langle\rangle.$$

As $\ddot{A}(a,z)$ only depends on $z$ and $e_c$ sends this variable to $c$, we also have $e_c[\ddot{A}(a,z)] = \ddot{A}(a,c)$. And, of course, $\mathbf{Wn}^{\ddot{A}(a,c)}\langle\rangle = \top$ means nothing but that $\ddot{A}(a,c)$ is true. So,
$$\mathbf{Wn}_{e_c}^{\bigsqcup x \ddot{A}(x,z)}\langle\top a\rangle = \top \text{ iff } \ddot{A}(a,c) \text{ is true}.$$

The above, together with (16), implies
$$\mathbf{Wn}^{\bigsqcup x \dot{A}^{\ddagger c}(x)}\langle\top a\rangle = \mathbf{Wn}_{e_c}^{\bigsqcup x \ddot{A}(x,z)}\langle\top a\rangle,$$

thus completing our proof of (15) and hence of (14).

Now, by induction, (14) extends from formulas of the form $\bigsqcup x \dot{A}(x)$ to all more complex subformulas of (10) including (10) itself, meaning that $(10)^{\ddagger c} = e_c[(10)^\star]$. The steps of this induction are straightforward, because each of the operations $\star$, $\ddagger c$ and $e_c[\ldots]$ commutes with each of the connectives $\wedge$, $\dot{\circ}$ and $\rightarrow$. $\square$

Putting (13) and Claim 7.8 together, we find that, for any constant $c$, $\Gamma_c$ is a $\bot$-won run of $e_c[(10)^\star]$. Now, remember that $\Gamma_c$ is the $\mathcal{H}_c$ vs. $\mathcal{E}$ run on $e_c$, according to Lemma 4.1 meaning that $\Gamma_c$ is the run spelled by an $e_c$-computation branch of $\mathcal{H}_c$. So, every $\mathcal{H}_c$ loses $(10)^\star$ on valuation $e_c$, i.e. no $\mathcal{H}_c$ wins $(10)^\star$. But every HPM is $\mathcal{H}_c$ for some $c$. Thus, no HPM wins $(10)^\star$. In other words, (10) is not valid.

This almost completes our proof of the main Claim 7.2 and hence our proof of the main Lemma 7.1 and hence our proof of the completeness of $\mathbf{Int}^{\circleddash}$. What remains to verify for the official completion of our proof of Theorem 3.1 for $\mathbf{Int}^{\circleddash}$ is that $^\star$ satisfies the complexity condition of Claim 7.2. Such a verification is given in the following subsection.

## 7.7 The complexity of the counterinterpretation $^\star$

The counterinterpretation $^\star$ constructed in the previous subsection interprets each atom $\dot{A}(x)$ of (10) as the binary predicate $\ddot{A}(x,z)$. With "true in the sense of Subsection 7.4 (resp. 7.5)" below understood as truth in the sense of the corresponding subsection where $B_c$ is taken in the role of $B$ (and, accordingly, $\Gamma_c$ in the role of $\Gamma$) when constructing the counterinterpretation $^\dagger$ from it, the meaning of the proposition $\ddot{A}(a,c)$ for any given constants $a,c$ is in fact the disjunction of the following two statements:

1. "$B_c$ is short and $\dot{A}(a)$ is true in the sense of Subsection 7.4";
2. "$B_c$ is long and $\dot{A}(a)$ is true in the sense of Subsection 7.5".

Note that arbitrarily long initial segments of $B_c$ can be effectively constructed from $c$. This can be done by first constructing the machine $\mathcal{H}_c$ from number $c$, and then tracing, step by step, how the play between $\mathcal{H}_c$ and $\mathcal{E}$ evolves on valuation $e_c$ according to the scenario described in the proof idea for Lemma 4.1. This makes it clear that the predicate "$B_c$ is long" (with $c$ here treated as a variable) is of complexity $\Sigma_1$, because it says nothing but that there is a computation step in $B_c$ at which $\lfloor \breve{W} \rfloor$ gets matchingly devirginized.

Next, "$B_c$ is short" is just the negation of "$B_c$ is long".

Next, some thought can show "$\dot{A}(a)$ is true in the sense of Subsection 7.4" (with $a$, together with the hidden $c$, here treated as a variable) to mean nothing but that there is no computation step in $B_c$ at which $\dot{A}(a)$ becomes the content of some positive non-matchingly devirginized residual molecule of the then-current position. So this is the negation of a $\Sigma_1$-predicate.

Finally, "$\dot{A}(a)$ is true in the sense of Subsection 7.5" means that there is a — lexicographically smallest — open chain $C$ whose last element is $\lfloor \breve{W} \rfloor$, and $\dot{A}(a)$ is not the content of anything in $C$. With a little thought we can see that, to find such a chain $C$, it would be sufficient to trace $B_c$ only up to the computation step at which $\lfloor \breve{W} \rfloor$ gets devirginized. So, the complexity of the predicate "$\dot{A}(a)$ is true in the sense of Subsection 7.5" is $\Sigma_1$.

To summarize, where $\dot{A}(x)$ is an atom of (10), $\ddot{A}(x,z)$, i.e $\dot{A}^\star(x)$, is indeed a Boolean combination of $\Sigma_1$-predicates. As for all other elementary atoms, we may make arbitrary assumptions — without affecting the incomputability of $(10)^\star$ — about how they are interpreted by $^\star$. So, we assume that they, too, are interpreted as Boolean combinations of $\Sigma_1$-predicates.

# 8  From ⚬− to ≻−

In this section we prove the remaining completeness part of Theorem 3.1 for **Int**$^\succ$. The following is an important lemma for our proof:

**Lemma 8.1** ⊩ ⚬$G \to \lambda G$, *for any formula $G$ of the language of affine logic.*

**Proof.**  The lemma will be proven by constructing an EPM $\mathcal{C}$ such that, for any static game $G$, $\mathcal{C} \models \, ⚬G \to \lambda G$. This, of course, immediately implies that we also have $\mathcal{C} ⊩ ⚬G \to \lambda G$ for any *formula* $G$.

Thinking of $\lambda G$ as the infinite conjunction $G \wedge G \wedge \ldots$, and thinking of any legal play over $⚬G \to \lambda G$ as consisting of two runs — one in the antecedent and one in the consequent, the idea of the work of $\mathcal{C}$, in intuitive terms, is the following. Every once in a while, in the antecedent, $\mathcal{C}$ makes a replicative move in the leaf $00\ldots0$ of the underlying bitstring tree. These are all replicative moves it makes, so that, in the eventual run $\Upsilon$ of $⚬G$, all of the strings $1, 01, 001, 0001, \ldots$ will be leaves of $Tree^{⚬G}\langle\Upsilon\rangle$ (and the only additional complete branch will be the infinite string of 0s). With each such leaf $0^m1$, $\mathcal{C}$ associates conjunct $\#m+1$ of $\lambda G$, and uses the copy-cat strategy between that conjunct and the copy of $G$ in leaf $0^m1$. There can be some delays here in mimicking adversary's moves, but this is OK as $G$ is static. Such a one-to-one matching guarantees that if $G$ is won by the adversary in all leaves of the bitstring tree generated in the antecedent, then all conjuncts of $\lambda G$ in the consequent are won by $\mathcal{C}$. In other words, if the adversary wins in $⚬G$, then $\mathcal{C}$ is sure to win in $\lambda G$. This translates into $\mathcal{C}$ being the winner in $⚬G \to \lambda G$.

In precise terms, here is the strategy that $\mathcal{C}$ follows, with $k$ and $\Phi$ being variables initialized 1 and $\langle\rangle$, respectively:

**PROCEDURE** LOOP: Let $w = 0^{k-1}$ (the string of $k-1$ zeros). Perform the following steps:

**Step 1.** Make the move $1.w$: .

**Step 2.** Where $(\Phi^{1.})^{\preceq w1} = \langle \bot\alpha_1, \ldots, \bot\alpha_m \rangle$ and $(\Phi^{2.})^{k.} = \langle \bot\beta_1, \ldots, \bot\beta_n \rangle$, make each of the moves  $2.k.\alpha_1$ , ..., $2.k.\alpha_m$ , $1.w1.\beta_1$ , ..., $1.w1.\beta_n$ .

**Step 3.** Grant permission. If the adversary responds by a move $\gamma$, then:

(i) if $\gamma = 2.j.\delta$ for some $1 \leq j \leq k$, then make the move $1.u.\delta$, where $u = 0^{j-1}1$ ;
(ii) if $\gamma = 1.v.\delta$ for some bitstring $v$, then, for each $j$ with $1 \leq j \leq k$ such that $v \preceq 0^{j-1}1$, make the move $2.j.\delta$ .

**Step 4.** Increment $k$ by one, update $\Phi$ to the current position of the play, and repeat LOOP.

Some additional informal explanations would help. In Step 1, $\mathcal{C}$ replicates leaf $0^{k-1}$ of (the underlying BT-structure of) the antecedent, thus creating two new leaves $0^k$ and $0^{k-1}1$. $0^k$ is just reserved for further replications, and otherwise $\mathcal{C}$ does not care about it. As for $0^{k-1}1$, it is "final" in that it will never be replicated in the future, thus forever remaining a leaf. To this leaf corresponds the $k$th conjunct of the consequent, in the sense that throughout the rest of the play, $\mathcal{C}$ will try, mimicking the adversary's moves, to keep the run of $G$ in the $k$th conjunct of the consequent identical with the run of $G$ in leaf $0^{k-1}1$ of the antecedent. Before this leaf was created, however, the adversary might have already made some moves $\beta_1, \ldots, \beta_n$ in the $k$th conjunct of the consequent, as well as some moves whose effect was making moves $\alpha_1, \ldots, \alpha_m$ in the (then-future) leaf $w1$ of the antecedent. $\mathcal{C}$ itself has not made any moves in those components during the previous iterations of LOOP, as at that time it did not treat $k$ as "active". Following the motto *better late than never*, during Step 2, $\mathcal{C}$ tries to catch up by copying the adversary's moves made in the $k$th conjunct of the consequent in the corresponding leaf $0^{k-1}1$ of the antecedent, and vice versa. This does not guarantee that the corresponding two runs of $G$ will be fully identical, but it does guarantee that, at least, one will be a $\top$-delay of the other, and this is just as good as if they were identical. During Step 3, for the sake of fairness, $\mathcal{C}$ lets the adversary make a move and, as long as such a move is a move in an "already activated" (i.e. $\leq k$th) conjunct of the consequent, or (the effect of the move is) in some already existing final leaves of the antecedent, the machine copies that move in the corresponding leaf of the antecedent or the corresponding conjuncts of the consequent.

In trying to verify that the above strategy is successful, as always, we may rely on the clean environment assumption, ruling out the possibility that the adversary ever makes illegal moves. It is not hard to see that then the machine does not make illegal moves either, so that we can narrow our attention down to legal runs only. Consider the run $\Gamma$ spelled by an arbitrary $e$-computation branch (whatever valuation $e$) of $\mathcal{C}$. Since LOOP is iterated infinitely many times and permission is granted once in each iteration, such a branch is fair. So, we just need to show that $\Gamma$ is a $\top$-won run of $e[\lozenge G \to \lambda G] = \lozenge e[G] \to \lambda e[G]$. For convenience, we may assume here that $G$ is a constant game so that $e[G] = G$, or otherwise just rename $e[G]$ into $G$. Thus, what we want to show is that $\mathbf{Wn}^{\lozenge G \to \lambda G}\langle\Gamma\rangle = \top$, which can be rephrased as $\mathbf{Wn}^{\wp \neg G \vee \lambda G}\langle\Gamma\rangle = \top$. As we agreed a while ago, here we can and do assume that $\Gamma$ is a legal run of $\lozenge G \to \lambda G$.

If $\mathbf{Wn}^{\lambda G}\langle\Gamma^{2.}\rangle = \top$, we are done. Suppose now $\mathbf{Wn}^{\lambda G}\langle\Gamma^{2.}\rangle = \bot$. This means that, for some $i$ ($i \geq 1$), $\Gamma^{2.i}$ is a $\bot$-won run of $G$. Hence $\neg\Gamma^{2.i}$ is a $\top$-won run of $\neg G$. Now, a little analysis of LOOP can convince us that, where $w = 0^{i-1}1$, $(\Gamma^{1.})^{\preceq w}$ is a $\top$-delay of $\neg\Gamma^{2.i}$. Therefore, as we deal with a static game, $(\Gamma^{1.})^{\preceq w}$, too, is a $\top$-won run of $\neg G$. This implies that $\mathbf{Wn}^{\wp \neg G}\langle\Gamma^{1.}\rangle = \top$. Hence $\mathbf{Wn}^{\wp \neg G \vee \lambda G}\langle\Gamma\rangle = \top$. Done. □

In Sections 5 and 6 we defined the notions of standardization and desequentization for formulas and sequents of the language of $\mathbf{Int}^{\circ\!\!-}$. The same concepts straightforwardly extend to $\mathbf{Int}^{\succ\!\!-}$-formulas and $\mathbf{Int}^{\succ\!\!-}$-sequents, with just $\circ\!\!-$ replaced by $\succ\!\!-$ and $\lozenge$ by $\lambda$ in the definitions.

**Lemma 8.2** *Assume $K$ is an $\mathbf{Int}^{\succ\!\!-}$-formula, and $D$ is the desequentization of the standardization of $K$. Then, for any interpretation $^*$ with $\models K^*$, we have $\models D^*$.*

**Proof.** This is an exact copy of Lemma 6.1, only with $\mathbf{Int}^{\succ\!\!-}$ instead of $\mathbf{Int}^{\circ\!\!-}$. Our proof of Lemma 6.1 relied on the fact that computability is closed under the rule

"*from $A$ to $\wedge A$*". The same closure principle holds with $\lambda$ instead of $\wedge$, as known from (the same) Section 13 of [10]. Everything else in the proof of Lemma 6.1 exclusively relied on the soundness of affine logic and certain syntactic facts about affine logic and implicative intuitionistic logic. But neither logic does syntactically discriminate between $\circ\!\!-$ and $\succ\!\!-$, or between $\wedge$ and $\lambda$. This means that our proof of Lemma 6.1 goes through for the present Lemma 8.2 virtually without any changes except just replacing $\wedge$ by $\lambda$ and $\circ\!\!-$ by $\succ\!\!-$ everywhere in the text of the proof. □

Now consider any $\mathbf{Int}^{\succ\!\!-}$-formula which is not provable in $\mathbf{Int}^{\succ\!\!-}$. Let us denote it by $K^{\succ\!\!-}$. Let $K^{\circ\!\!-}$ be the same formula, only with every occurrence of $\succ\!\!-$ replaced by $\circ\!\!-$. Next, let $\mathcal{S}^{\circ\!\!-}$ and $\mathcal{S}^{\succ\!\!-}$ be the standardizations of $K^{\circ\!\!-}$ and $K^{\succ\!\!-}$, respectively. We may assume that the formula (9) from the previous section is the desequentization of $\mathcal{S}^{\circ\!\!-}$, and hence the following formula is the desequentization of $\mathcal{S}^{\succ\!\!-}$:

$$\begin{array}{c} \lambda(X_1 \wedge Y_1 \to Z_1) \wedge \cdots \wedge \lambda(X_s \wedge Y_s \to Z_s) \wedge \\ \lambda\big((\lambda P_1 \to Q_1) \to R_1\big) \wedge \cdots \wedge \lambda\big((\lambda P_s \to Q_s) \to R_s\big) \end{array} \quad \to \quad W. \qquad (17)$$

The fact that $\mathbf{Int}^{\succ\!\!-} \not\vdash K^{\succ\!\!-}$, of course, implies $\mathbf{Int}^{\circ\!\!-} \not\vdash K^{\circ\!\!-}$. This, in turn, by Lemma 5.2, implies $\mathbf{Int}^{\circ\!\!-} \not\vdash \mathcal{S}^{\circ\!\!-}$. Therefore, by Lemma 7.1, there is an interpretation $*$ satisfying the complexity condition of clause (b) of Theorem 3.1 such that $\not\models (9)^*$. Note that (17) can be obtained from (9) through repeatedly — $3s$ times — replacing a negative occurrence of a subformula $\wedge G$ by $\lambda G$. Lemmas 2.1(b) and 8.1 then guarantee that $\Vdash (17) \to (9)$, which implies $\models (17)^* \to (9)^*$. From here we infer that, if $\models (17)^*$, then $\models (9)^*$, because modus ponens preserves computability. But $\not\models (9)^*$, and thus $\not\models (17)^*$. Therefore, by Lemma 8.2, $\not\models (K^{\succ\!\!-})^*$. Our completeness proof for $\mathbf{Int}^{\succ\!\!-}$ is hereby complete, and the main Theorem 3.1 of this paper is now fully proven.